\documentclass[a4paper,11pt]{article}
\usepackage{jcappub} 

\arxivnumber{2504.12105} 
\usepackage{graphicx}

\usepackage[normalem]{ulem}

\usepackage{float}
\usepackage{amsmath}
\usepackage{amssymb}
\usepackage{booktabs}
\usepackage{hhline}
\usepackage{multirow}
\usepackage[rightcaption]{sidecap}
\usepackage{hyperref}
\usepackage{xcolor}
\usepackage{wrapfig}
\usepackage{braket}

\graphicspath{ {./Figures/} }

\def\beq{\begin{equation}}
\def\eeq{\end{equation}}
\def\bea{\begin{eqnarray}}
\def\eea{\end{eqnarray}}
\def\l{\left}
\def\r{\right}
\def\ddd{\mathrm{d}}
\def\nn{\nonumber}

\usepackage{lineno}

\title{\boldmath Can asteroid-mass PBHDM be compatible with catalyzed phase transition interpretation of PTA?}

\author[a,b]{Jiahang Zhong}
\emailAdd{jiahangzhong@mail.ustc.edu.cn}
\author[c]{Chao Chen}
\emailAdd{cchao012@just.edu.cn}
\author[a,b]{Yi-Fu Cai}
\emailAdd{yifucai@ustc.edu.cn}

\affiliation[a]{Department of Astronomy, School of Physical Sciences, \\
University of Science and Technology of China, Hefei 230026, China}

\affiliation[b]{
CAS Key Laboratory for Research in Galaxies and Cosmology,\\
School of Astronomy and Space Science, University of Science and Technology of China,\\
Hefei 230026, China}

\affiliation[c]{
School of Science, Jiangsu University of Science and Technology, \\
Zhenjiang, 212100, China}

\abstract{Primordial black holes (PBHs) can catalyze first-order phase transitions (FOPTs) in their vicinity, potentially modifying the gravitational wave (GW) signals from PTs. 
In this study, we investigate the GWs from strong PTs catalyzed by PBHs. We consider high PBH number densities, corresponding to asteroid-mass PBH dark matter (DM) when the GWs from FOPTs peak in the nanohertz band. 
We calculate the PBH-catalyzed FOPT GWs from both bubble collision GWs and scaler-induced gravitational waves (SIGWs). We find that while low PBH number densities amplify the GW signals due to the formation of large bubbles, high PBH number densities suppress them, as the accelerated phase transition proceeds too rapidly. This suppression renders the signals unable to explain pulsar timing array (PTA) observations.  
By conducting data fitting with the NANOGrav 15-year dataset, we find that the PBH catalytic effect significantly alters the estimation of PT parameters. 
Notably, our analysis of the bubble collision GWs reveals that, the asteroid-mass PBHs ($10^{-16}-10^{-12} M_\odot$) constituting all DM is incompatible with the PT interpretation of PTA signals. 
However, incorporating SIGWs alleviates this incompatibility for PBHs in the mass range $10^{-14} - 10^{-12} M_\odot$. 
}

\begin{document}

\maketitle
\flushbottom


\section{Introduction}
\label{sec:intro}


First-order phase transitions (FOPTs) are predicted by new physics beyond the standard model~\cite{EWPT1,PT2,PT3,PT4} for various motivations, such as dark matter production~\cite{Falkowski_2013,Cai:2024nln,lewicki_black_2024,franciolini_curvature_2025}, baryogenesis~\cite{Morrissey:2012db}, and the generation of a stochastic gravitational wave background (SGWB)~\cite{Witten:1984rs,Hogan:1986dsh,Athron:2023xlk}. The relevant SGWB signal is expected to be probed by upcoming gravitational wave (GW) experiments such as LISA~\cite{LISA:2017pwj}, DECIGO~\cite{DECIGO1}, BBO~\cite{BBO}, Taiji~\cite{Taiji} and TianQin~\cite{TianQin:2015yph}, offering a unique window into early-universe PTs and new physics.  

{
Recent pulsar timing array (PTA) observations reported a stochastic gravitational wave background (SGWB)~\cite{NANOGrav:2023gor,InternationalPulsarTimingArray:2023mzf,EPTA:2023fyk,Xu:2023wog,Reardon:2023gzh}. 
While the signal is broadly compatible with GWs from the mergers of supermassive black holes~\cite{NANOGrav:SMBH,Ellis:2023dgf}, reconciling its amplitude with prior merger-density estimates remains problematic~\cite{Casey-Clyde:2021xro,Kelley:2016gse,Kelley:2017lek,Shen:2023pan}, leaving the astrophysical origin an open issue. 
A compelling possibility is that SGWB arises from cosmological sources~\cite{NANOGrav:2023hvm}, most notably from FOPTs~\cite{Nakai:2020oit,Ratzinger:2020koh,Xue:2021gyq,Deng:2023twb,Megias:2023kiy}.
Given the large amplitude of the SGWB, strong phase transitions are usually assumed to accommodate this feature naturally. In such case, the energy of the transition dominates the Hubble expansion and the main contributions of the GWs are the bubble collision GWs and accompanied SIGWs.\footnote{FOPTs are naturally accompanied by SIGWs since curvature perturbation would be generated from the asynchronous vacuum decay~\cite{Lewicki:2022pdb}. However, the result of the PT accompanied SIGWs remains a topic of debate~\cite{lewicki_black_2024,lewicki_black_2024a,franciolini_curvature_2025,zou_numerical_2025}.} 
A strong FOPT near \(T \sim0.1\ {\rm GeV}\) can explain the PTA signal~\cite{ellis_what_2023}.

Primordial black holes (PBHs)~\cite{Zeldovich:1967lct,Hawking:1971ei,Carr:1974nx}, which have been associated with numerous astrophysical phenomena, including serving as candidates for dark matter~\cite{Carr:2020gox,PBHdarkmatter,Khlopov:2008qy}, explaining the formation of supermassive black holes at galactic nuclei~\cite{nuclei,nuclei2}, and the binary black hole merger events detected by LIGO/Virgo~\cite{Raidal:2017mfl}, can be generated from large fluctuations~\cite{sasaki_primordial_2018,Cai:2024nln} in the very early universe. The masses of PBHs are related to the Hubble horizon at the formation time~\cite{sasaki_primordial_2018}, \(M_{\rm pbh} =\gamma\frac{1}{2G} H^{-1}_{\rm form}\), where \(\gamma\sim 0.2\) is a correction factor.
Recent studies highlight that PBHs can catalyze FOPTs~\cite{Burda2015VacuumMetastabilitywithBlackHoles,Canko2018OntheCatalysisoftheElectroweakVacuumDecaybyBlackHolesatHighTemperature,Gregory2014BlackHolesAsBubbleNucleationSites,Gregory2016TheFateoftheHiggsVacuum,Gregory2020BlackHolesOscillatingInstantonsandtheHawkingMossTransition,Hiscock1987CanBlackHolesNucleateVacuumPhaseTransitions,Kohri2018ElectroweakVacuumCollapseInducedbyVacuumFluctuationsoftheHiggsFieldaroundEvaporatingBlackHoles,Moss1985BlackHoleBubbles,Mukaida2017FalseVacuumDecayCatalyzedbyBlackHoles,Strumia2023BlackHolesDon’tSourceFastHiggsVacuumDecay,oshita_compact_2019}, if they formed before the FOPTs. Importantly, PBHs can initiate local PTs before background nucleation rate achieves nucleation condition, leading to a accelerated PT and alteration of the spacetime-distribution of bubble nucleation. The alteration of spacetime-distribution in bubble nucleation inevitably induces systematic modifications to the GW spectrum~\cite{jinno_superslow_2024,zeng_phase_2024}.

PBHs with masses lower than a solar mass formed before the universe cooled to temperatures of $T \sim 0.1 \ \mathrm{GeV}$~\cite{Franciolini:2023pbf}, and their catalytic effect can impact the PT interpretation of PTA signals. 
As we will discuss in Sec.~\ref{sec: PBHinfluence}, if PBHs constitute all DM with asteroid masses \(10^{-16}M_{\odot}\sim 10^{-11}M_{\rm \odot}\), they correspond to high number densities (exceeding one PBH per Hubble horizon) around $T \sim 0.1 \ \mathrm{GeV}$, amplifying their impact on the PT dynamics. Prior investigations~\cite{jinno_superslow_2024,zeng_phase_2024} have studied catalyzed PTs with only low PBH number densities, uncovering interesting phenomena. However, these analyses omitted the high-density PBH scenario and thus cannot address the situation where asteroid-mass PBH constituting all DM affect PTs at around \(T\sim 0.1\ {\rm GeV}\). Our work therefore establishes a unified framework to characterize the catalytic effects of PBHs—spanning low to high number densities—on strong FOPT GWs. Using this framework, we then examine how PBHs modify the PT interpretation of PTA signals. Specifically, we investigate whether the asteroid-mass PBH dark matter hypothesis conflicts with the cosmological phase transition explanation of PTA signals.
}
For simplicity, in this study, we assume a monochromatic PBH mass function and a random spatial distribution of PBHs without initial clustering~\cite{ding_detectability_2019}.

The organization of the paper is as follows. In Sec.~\ref{sec: PBHinfluence}, we present the basic setup for the description of phenomenological PT models and the bubble nucleation rate in the presence of PBHs. In Sec.~\ref{sec:PTGW}, we discuss how to estimate GWs from PBH-catalyzed PTs. After that, we calculate the GW spectral shape for the bubble collision GWs in Sec.~\ref{subsec: PGW} and the SIGWs generated during PTs in Sec.~\ref{subsec: SIGW}. In Sec.~\ref{sec:influence}, we perform data fitting with the NANOGrav 15-year dataset and finally discuss our results in Sec.~\ref{sec:disc}.

\section{PBH Catalytic Effect}
\label{sec: PBHinfluence}

We first simply review the phenomenological model of very strong PTs in the early universe without PBHs existence. The general form of bubble nucleation rate per spacetime can be expressed as~\cite{Ellis:2018mja,Coleman:1977py,Enqvist:1991xw} 
\begin{equation}
    \Gamma_0(t) = \Gamma_0(t_\star)e^{\beta(t-t_\star)} ~,
\end{equation}
where, \(t_\star\) is the cosmic time near PT, \(\beta\) is the inverse duration of PT and \(\Gamma_0(t_\star)\) is the nucleation rate at a given time. Both \(\beta\) and \(\Gamma_0(t_\star)\) can be calculated from a given particle physics model~\cite{Linde:1981zj,Coleman:1977py},
\begin{equation}
    \beta = HT_\star \frac{d(S_3/T)}{dT} \Big|_{T=T_\star}~,\quad \Gamma_0(t_\star) \sim T_\star^4\l(\frac{S_3/T_\star}{2\pi}\r)^{\frac{3}{2}} ~,
    \label{eq:PTmodel}
\end{equation}
where \(S_3\) is the three-dimensional bounce action. We can estimate the PT nucleation time \(t_n\) as \(\Gamma_0(t_n) \sim H^4\), where \(H^2 = \rho_V/M_{\rm pl}^2\) and \(\rho_V\) is vacuum energy density given by particle physics models, which is equal to total energy density in the universe since we assume a very strong PT. After PT completes, plasma will be reheated\footnote{This reheating, induced by the latent heat of PTs, should not be confused with the reheating following cosmic inflation.} to temperature \(T_{\rm re} = \l(\frac{90 \rho_V}{\pi^2g_\star}\r)^{1/4}\). 

As discussed in Refs.~\cite{zeng_phase_2024,Mukaida2017FalseVacuumDecayCatalyzedbyBlackHoles,Hiscock1987CanBlackHolesNucleateVacuumPhaseTransitions}, PBHs with mass smaller than \(M_{\rm pl}^3/\sqrt{V_{f}}\) can catalyze FOPTs, where \(M_{\rm pl} \equiv 1/\sqrt{8\pi G} \) is the reduced Plank mass and \(V_{f}\) is the energy density of the false vacuum. This estimation ignores the change of the Bekenstein entropy, which only considers the pure gravitational effect. For a FOPT with \(T_{\rm re}\sim 0.1{\rm GeV}\), the energy density of the false vacuum is \(V_{f} \sim 10^{-4}{\rm GeV}^4\), which suggests that PBHs with mass less than \(10^{7}M_\odot\) can catalyze the PT. When such PBHs exist in the early universe, they act as localized catalysts that nucleate bubbles encapsulating them, triggering local phase transitions much earlier than the background would. Here we consider PBHs with masses less than a solar mass since they formed before \(T\sim 0.1\ {\rm GeV}\). The nucleation rate is expressed as 
\begin{align}
    \Gamma(t) = \Gamma_0(t) + \Gamma_\mathrm{pbh}(t) = \Gamma_0(t_\star)e^{\beta (t-t_\star)}+ \Gamma_\mathrm{pbh}(t)~,
    \label{eq:nucleationrate}
\end{align}
where \(\Gamma_\mathrm{pbh}\) denotes the spatially averaged nucleation rate from PBH catalytic effect. This catalytic effect only relies on the masses of PBHs in a given PT~\cite{Hiscock1987CanBlackHolesNucleateVacuumPhaseTransitions,Mukaida2017FalseVacuumDecayCatalyzedbyBlackHoles}. Here, we take the initial mass of PBHs, $M_{PBH,i}$, to be monochromatic so that bubbles nucleate simultaneously around all PBHs~\cite{jinno_superslow_2024}. Therefore, after spatial averaging, the nucleation rate can be estimated as~\cite{jinno_superslow_2024,zeng_phase_2024}
\begin{align}
\Gamma_\mathrm{pbh} = n_\mathrm{pbh}(t_c) H^3\delta(t-t_c) ~, 
\end{align}
where \(n_\mathrm{pbh}(t_c)\) is normalized PBH number density which denotes averaged PBH number per unit Hubble volume at time \(t_c\). The difference between \(t_c\) and \(t_n\) reflects the catalytic strength of PBHs. We thus define
\begin{equation}
    G_0 = \frac{\Gamma_0(t_c)}{H^4}\ll 1~,
\end{equation}
as the catalytic strength. Note that \(\Gamma/H^4\sim 1\) indicates the beginning of a normal PT. Depending on PBH masses and PT models, PBHs can reduce the tunneling action by a factor of \(\mathcal{O}(1\sim 10)\)~\cite{Hiscock1987CanBlackHolesNucleateVacuumPhaseTransitions,Mukaida2017FalseVacuumDecayCatalyzedbyBlackHoles,Canko2018OntheCatalysisoftheElectroweakVacuumDecaybyBlackHolesatHighTemperature,Gregory2014BlackHolesAsBubbleNucleationSites,Gregory2016TheFateoftheHiggsVacuum}. This corresponds to \(G_0\approx 10^{-10}\sim 10^{-100}\). Here, we treat the catalytic strength \(G_0\) as a free parameter, without specifying a PT model. We adopt \(G_0\approx 10^{-10}\) as a conservative estimation.

The normalized PBH number density is given by
\begin{align}
n_\mathrm{pbh}(t) H^3 = \l(\frac{a(t)}{a_0}\r)^{-3} \rho_{c,0}\Omega_{\rm DM,0}\frac{f_{\rm pbh}}{M_{\rm pbh}} ~.
\end{align}
where \(f_{\rm pbh}\) is the present PBH mass fraction, \(\rho_{c,0}=3H_0^2/(8\pi G)\) is the current critical energy density and \(\Omega_{\rm DM,0}\) is the current normalized dark matter density. The current normalized PBH number density can be estimated as
\begin{align}
n_{\rm pbh}(t_0) H_0^3 \approx 3.438\times 10^{10} \l(\frac{M_\odot}{M_{\rm pbh}}\r)\l(\frac{f_{\rm pbh}}{1.0}\r) {\rm Mpc}^{-3}~.
\end{align}
Using entropy conservation\footnote{Here we adopt the effective number of neutrino species \(N_{\rm eff} = 3\), given that \(N_{\rm eff} = 3.02 \pm 0.17\) from Planck 2018~\cite{Planck:2018jri}.}
\begin{align}
    g_* a^3(t_{\rm re})T_{\rm re}^3 = \frac{43}{11}a_0^3T_0^3~, 
\end{align}
with the reheating temperature \(T_{\rm re}=0.1\) GeV, we obtain,
\begin{equation}
    n_{\rm pbh}(t_{\rm re}) \approx 1.355\times 10^{-8} \l(\frac{M_\odot}{M_{\rm pbh}}\r)\l(\frac{f_{\rm pbh}}{1.0}\r)\l(\frac{0.1\ \rm GeV}{T_{\rm re}}\r)^3 \l(\frac{g_*}{100}\r)^{-1/2}~,
    \label{eq:pbhdensity}
\end{equation}
where \(t_{\rm re}\) is the corresponding time of the reheating after a FOPT. For simplicity, we ignore the cosmic expansion during the PT in the following discussions, and thus $n_\mathrm{pbh}(t_c) = n_\mathrm{pbh}(t_{\rm re})$. However, this assumption may modify our results if the PT is super-slow, and we leave this to the future study. Notice that when \(T_{\rm re}\sim 0.1 {\rm GeV}\) and \(f_{\rm pbh} = 1\), PBHs with \(M_{\rm pbh}\sim 10^{-15}M_\odot\) gives \(n_{\rm pbh} = 10^7\), which is really dense. Even a small mass fraction \(f_{\rm pbh}\sim 10^{-5}\) will also correspond to high number density \(n_{\rm pbh}\sim 10^{2}\).


We now look at how PBHs affect the spatial-averaged, nucleation-time-distribution of bubbles. For a bubble to nucleate at a given four-dimensional point $\left(t_n, \vec{x}_n\right)$ with an infinitesimal spacetime volume element $\ddd^4 x_n= \ddd t_n \ddd^3 x_n$, it needs to satisfy two conditions~\cite{jinno_effect_2021}: (1) No bubble nucleates inside the past light cone of $\left(t_n, \vec{x}_n\right)$. (2) One bubble nucleates in $\ddd^4 x_n$.
The former probability, which we call survival probability $P_{\text {surv }}\left(t_n, \vec{x}_n\right)$, can be expressed as
\begin{align}
P_{\text {surv }}\left(t_n, \vec{x}_n\right) & =\prod_{x_c \in \text { past light cone of }\left(t_n, \vec{x}_n\right)}\left[1-\Gamma\left(x_c\right) \ddd^4 x_c\right] \notag
\\
& =\exp \left[-\int_{x_c \in \text { past light cone of }\left(t_n, \vec{x}_n\right)} \ddd^4 x_c \Gamma\left(x_c\right)\right] \notag
\\
& =\exp\left[-8 \pi \frac{\Gamma_0(t_\star)}{\beta^4} e^{\beta (t_n-t_\star)}-\frac{4\pi}{3} n_\mathrm{pbh} H^3 (t_n-t_c)^3\right]~.
\label{eq:surv}
\end{align}

In the last equality, we have neglected the effect of cosmic expansion. Here, for simplicity, we have denoted \(n_{\rm pbh}(t_{c})\) as \(n_{\rm pbh}\). Together with the latter probability, which is equal to the nucleation rate $\Gamma(t)$, the nucleation-time-distribution $P_{\rm nuc }\left(t_n\right)$ is
\begin{align}
P_{\rm nuc}\left(t_n\right) A= &  \Gamma_0(t_\star) \exp\left[\beta (t_n-t_\star)-8 \pi \frac{\Gamma_0(t_\star)}{\beta^4} e^{\beta (t_n-t_\star)}-\frac{4\pi}{3} n_\mathrm{pbh} H^3 (t_n-t_c)^3\right] \nn\\
&+n_\mathrm{pbh} H^3 \delta(t_n-t_c) \exp\left[-8 \pi \frac{\Gamma_0(t_c)}{\beta^4} \right]~.
\label{eq:distribute}
\end{align}

 Note that the overall factor \(A\) is chosen so that $\int_{t_c}^{\infty} d t_n P_{\rm nuc}\left(t_n\right)=1$. We can further simplify Eq.~\eqref{eq:distribute} by choosing $t_\star=t_c=0$ and using the fact that $\Gamma_0(t_c)/\beta^4 \ll 1$, which gives

\begin{align}
P_{\rm nuc}\left(t_n\right) A/\beta^4=   \frac{\Gamma_0(\Tilde{t}_c)}{\beta^4} \exp\left[\Tilde{t}_n -8 \pi \frac{\Gamma_0(\Tilde{t}_c)}{\beta^4} e^{\Tilde{t}_n }-\frac{4\pi}{3} n_\mathrm{pbh}\frac{H^3}{\beta^3} \Tilde{t}_n^3\right] +n_\mathrm{pbh}\frac{H^3}{\beta^3}\delta(\Tilde{t}_n)~,
\label{eq:distribute2}
\end{align}
where $\Tilde{t}= \beta t$, $\Gamma_0/\beta^4$ and $n_\mathrm{pbh} H^3/\beta^3$ are dimensionless variables rescaled by \(\beta\).

\begin{figure}[!tb]
\centering
\includegraphics[width=0.45\textwidth]{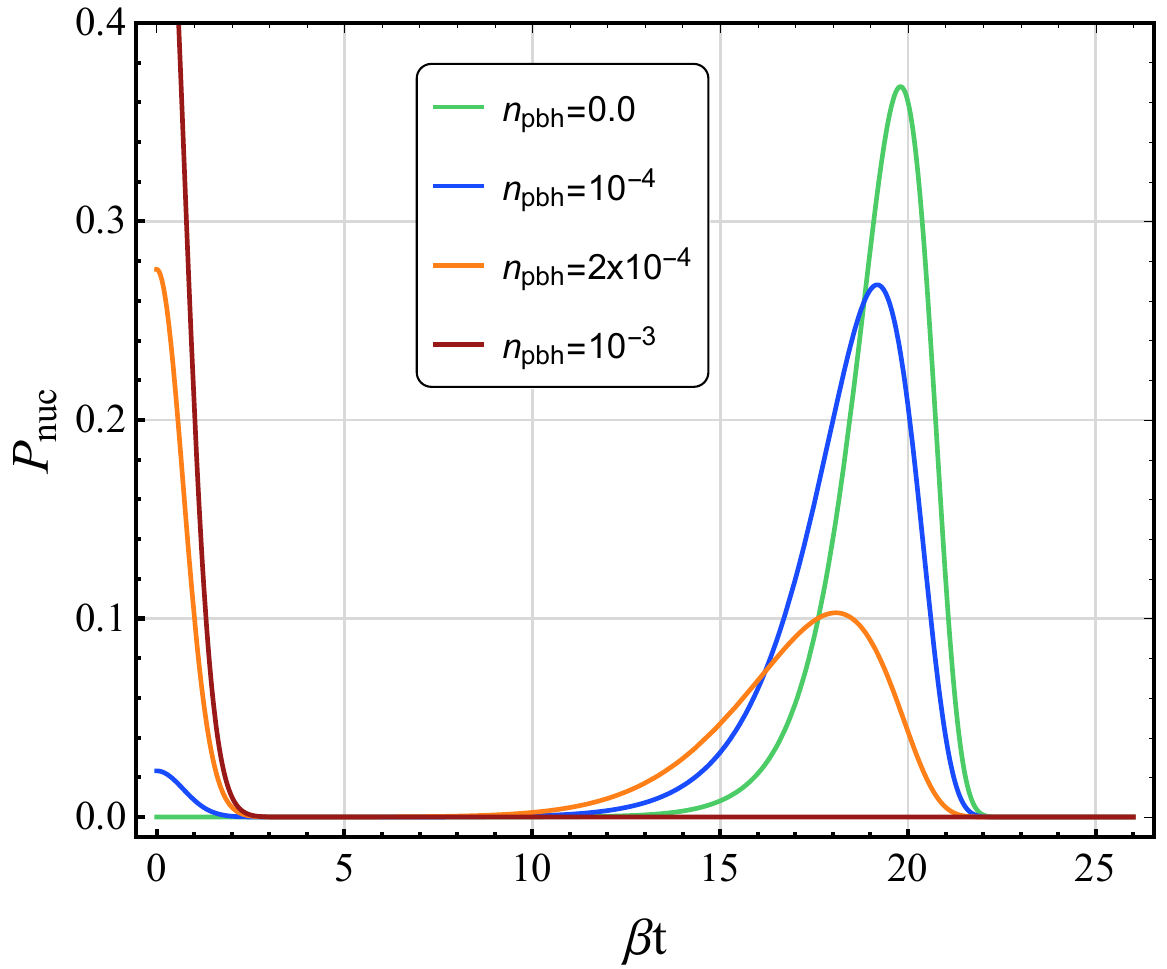}
\includegraphics[width=0.45\textwidth]{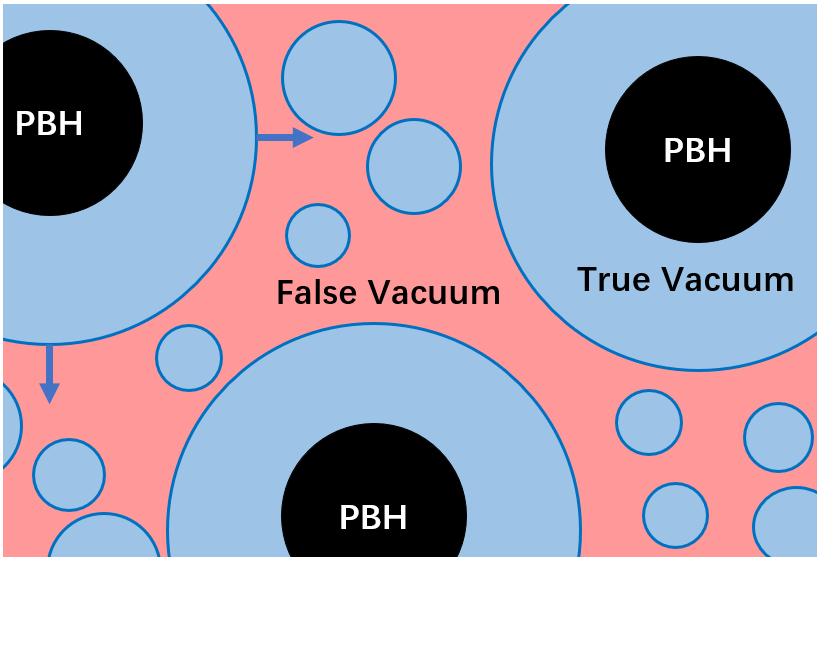}
\caption{{\it Left:} The nucleation-time-distributions with various given normalized PBH number densities. We have chosen \(G_0= 10^{-10}\) and \(\beta/H = 1\). For visibility, we plot DiracDelta function by using approximate Guassian function expression. {\it Right:} Schematic diagram of PBH-catalyzed PTs. Universe transfered from unstable false vacuum to true vacuum.}
\label{fig:nucleationtime}
\end{figure}

We can see from the Left panel of Fig.~\ref{fig:nucleationtime} that the catalytic effect of PBHs can significantly change the bubble nucleation-time distribution. Without PBHs, the bubble nucleation-time distribution first increases due to the exponential growth of the nucleation rate and then reaches a peak value due to suppression caused by the decreasing survival possibility. However, if there are PBHs in the universe before the PT, bubbles would nucleate near PBHs before the universe drops to nucleation temperature, forming one new peak for the bubble nucleation-time distribution. For those causal regions without PBHs, bubbles nucleate as in normal PTs, forming another peak for the bubble nucleation-time distribution in later time, which is suppressed since bubbles catalyzed by PBHs will reduce late-time survival possibilities. The Right panel of Fig.~\ref{fig:nucleationtime} shows a schematic diagram of the spatial distribution of bubbles. Note that there are larger bubbles existing around PBHs since bubbles nucleated earlier around PBHs than background. This unique nucleation spacetime distribution of  bubbles can modify the GW signals, allowing us to identify PBH influence through GW observations. In the following, we will discuss the PBH influence on bubble collision GWs and SIGWs. 

\section{GWs from Catalyzed PTs}
\label{sec:PTGW}
{
We first briefly review the approximation of the GW spectrum derived with dimensional analysis in the absence of PBHs, following arguments similar to those in ref.~\cite{Kosowsky:1991ua,Athron:2023xlk}. We assume that the energy in the GWs must be proportional to Newton’s constant, \(G\). Furthermore, the energy is assumed to depend on the available vacuum energy, \(\kappa\rho_V\) , where \(\kappa\) denotes the fraction of vacuum energy available for GWs, the bubble wall velocity \(v_w\), and that the only other relevant dimensionful scale is the characteristic timescale \(\beta_e^{-1}\) (not necessarily that derived from Eq.~\eqref{eq:PTmodel}). On dimensional grounds, we must have that
\begin{align}
    E_{\rm GW} \sim Gv_w^3\kappa^2\rho_V^2\beta_e^{-5}~.
\end{align}
In the very strong phase transitions, the bubble wall velocity approaches light speed \(v_w\sim 1\). The total liberated vacuum energy, on the other hand, is not proportional to G and on similar dimensional grounds, we must have that
\begin{align}
    E_V \sim \rho_V v_w^3\beta_e^{-3}~.
\end{align}
Thus, using $H \sim \sqrt{G\rho_{\mathrm{tot}}}$ and defining $\alpha = \rho_V / \rho_R$ such that $\rho_V / \rho_{\mathrm{tot}} = \alpha / (1 + \alpha)$, the normalized energy density of GWs can be written as
\begin{align}
\frac{\rho_{\mathrm{GW}}}{\rho_{\mathrm{tot}}} \sim \frac{\rho_V}{\rho_{\mathrm{tot}}} \frac{E_{\mathrm{GW}}}{E_V} \sim \kappa^2 \left(\frac{\alpha}{1 + \alpha}\right)^2 \left(\frac{\beta_e}{H}\right)^{-2}~.
\end{align} 
Finally, the spectrum can then be written as
\begin{align}
\Omega_{\rm GW} =\frac{1}{\rho_{\mathrm{tot}}} \frac{d\rho_{\mathrm{GW}}}{d\ln f} \sim \kappa^2 \left(\frac{\alpha}{1 + \alpha}\right)^2 \left(\frac{\beta_e}{H}\right)^{-2} \Delta(f\beta_e^{-1})~,
\label{eq:gwform}
\end{align}  
where the function $\Delta$ was forced to be a function of $f\beta_e^{-1}$ and we expect a peak frequency at around $\beta_e$ because $\beta_e^{-1}$ remains the only relevant timescale in the problem. We assume \(\alpha\gg1\) and \(\kappa=1\) since we consider strong PT scenarios. It is notable that in our PBH-catalyzed PT scenarios, PBHs alter the GW spectrum only through their ability to change the bubble distribution. Therefore we expect that PBHs will not affect the bubble wall velocity \(v_w\), the vacuum energy density \(\rho_V\) and the fraction \(\kappa\) the influence of PBHs will only be presented in relevant timescale \(\beta_e(\beta,n_{\rm pbh})\), which is related to mean bubble separation as \(\beta_e^{-1} \sim R_\star/v_w\sim R_\star\). With the existence of PBHs, the mean bubble separation can be calculated from the bubble distribution,
\begin{align}
R_\star &= \l(n_{\rm bubble}\r)^{-1/3} = \l(\int_{t_c}^{t_p} \ddd t\Gamma(t)F(t)\r)^{-1/3},
\\
F(t)&= \exp\l( -\int_{t_c}^{t}\ddd t' \frac{4\pi}{3}\Gamma(t') r^3(t',t)\r),
\end{align}
where \(F(t)\) is the averaged false vacuum fraction~\cite{Turner:1992tz} and \(r(t',t) = t'-t\) when ignoring the cosmic expansion. Here we use the value of mean bubble separation at percolation time~\cite{Lin2018ContinuumPO,Li2020NumericalSF}, solved by \(F(t_p)\approx 0.71\), which denotes the onset of bubble collisions. We will see later that this choice yields excellent agreement with the numerical results obtained using envelope approximation. Note that the function \(\Gamma(t)\times F(t)\) is numerically equal to \(\ P_{\rm nuc}(t) A\) defined in Eq.~\eqref{eq:distribute}. We can directly get 
\begin{align}
R_\star^{-3} = n_\mathrm{pbh} H^3+\beta^3\int_{0}^{\Tilde{t}_p} \ddd \Tilde{t} \l(\frac{\Gamma_0(\Tilde{t}_c)}{\beta^4} \exp\left[\Tilde{t}_n -8 \pi \frac{\Gamma_0(\Tilde{t}_c)}{\beta^4} e^{\Tilde{t}_n }-\frac{4\pi}{3} n_\mathrm{pbh}\frac{H^3}{\beta^3} \Tilde{t}_n^3\right] \r)~.
\label{eq:Rsep}
\end{align}
Thus the effective inverse timescale is 
\begin{align}
\frac{\beta_e}{\beta} = \frac{R_\star(n_{\rm pbh}=0)}{R_\star(n_{\rm pbh})} ~.
\label{eq:betaeff}
\end{align}
From Eq.~\eqref{eq:gwform}, the peak frequencies \(f_p\) and peak value of spectrum \(\Omega_p\) can be expressed as
\begin{align}
f_p(n_{\rm pbh}) \approx f_p(n_{\rm pbh} = 0) \beta_e/\beta ,\quad \Omega_p(n_{\rm pbh}) \approx \Omega_p(n_{\rm pbh} = 0) \beta^2/\beta_e^{2}~, 
\end{align} 
The Left panel of Fig.~\ref{fig:betafit} shows the the analytical results of \(\beta_e/\beta\) in Eq.~\eqref{eq:betaeff}, along with the numerical results of \(f_p(n_{\rm pbh})/f_p(n_{\rm pbh}=0)\), \(\sqrt{\Omega_p(n_{\rm pbh}=0)/\Omega_p(n_{\rm pbh})}\) obtained using the envelope approximation as we will discuss later. When the PBH number densities are relative smalle, they can provide large bubbles due to their catalytic effect, which leads to larger mean bubble separation, i.e. \(\beta_e/\beta<1\), thereby amplifying GWs emission. However, relative high PBH number densities will quickly drive \(\beta_e/\beta>1\) since a large amount of bubbles were generated at the beginning, which leads to suppressed GW signals. Note that when PT inverse duration \(\beta\) is small compared to normalized PBH number density, i.e., PTs are dominated by PBHs catalytic effect, We can neglect the first term in Eq.~\eqref{eq:Rsep} and the effective \(\beta_e\) can be derived as \( \beta_e/H \approx 4.37\,n_{\rm pbh}^{1/3}\). The threshold of \(\beta_e/\beta>1\) can be approximately solved from \(4.37\,n_{\rm pbh}^{1/3} =\beta/H\), which gives
\begin{align}
    n_{\rm pbh,c} \frac{H^3}{\beta^3} \approx 0.01~.
\end{align}
The \(n_{\rm pbh,c}\) represents the characteristic value of PBH number densities. When \(n_{\rm pbh}\gg n_{\rm pbh,c}\), i.e. in PBH-dominated region, the inverse timescale \(\beta_e\) is an increasing function of \(n_{\rm pbh}\), while there is a minimum around \(n_{\rm pbh}H^3/\beta^3\sim 10^{-4}\) appear in the background-dominated region \(n_{\rm pbh}\ll n_{\rm pbh,c}\). It is notable that PBH-dominated PT does not necessarily mean high PBH number density, which stands for \(n_{\rm pbh}>1\). 
The Right panel of Fig.~\ref{fig:betafit} shows analytical values of \(\beta_e/\beta\) in plane (\(n_{\rm pbh}\), \(\beta\)). With these analysis, we can see the parameter space for prior study~\cite{jinno_superslow_2024,zeng_phase_2024}. Ref.~\cite{jinno_superslow_2024}\footnote{They, instead, discussed PBH's catalytic effect in weak PT. } set \(\beta/H\sim 0\) and thus they studied PBH-dominated PT but with low PBH number densities \(n_{\rm pbh}<1\), while Ref.~\cite{zeng_phase_2024} studied background-dominated PT with low PBH number densities. In this study, to address the hypothesis of asteroid-mass PBH as whole dark matter, we will focus on high PBH number densities.

\begin{figure}[!tb]
\centering
\includegraphics[width=0.45\textwidth]{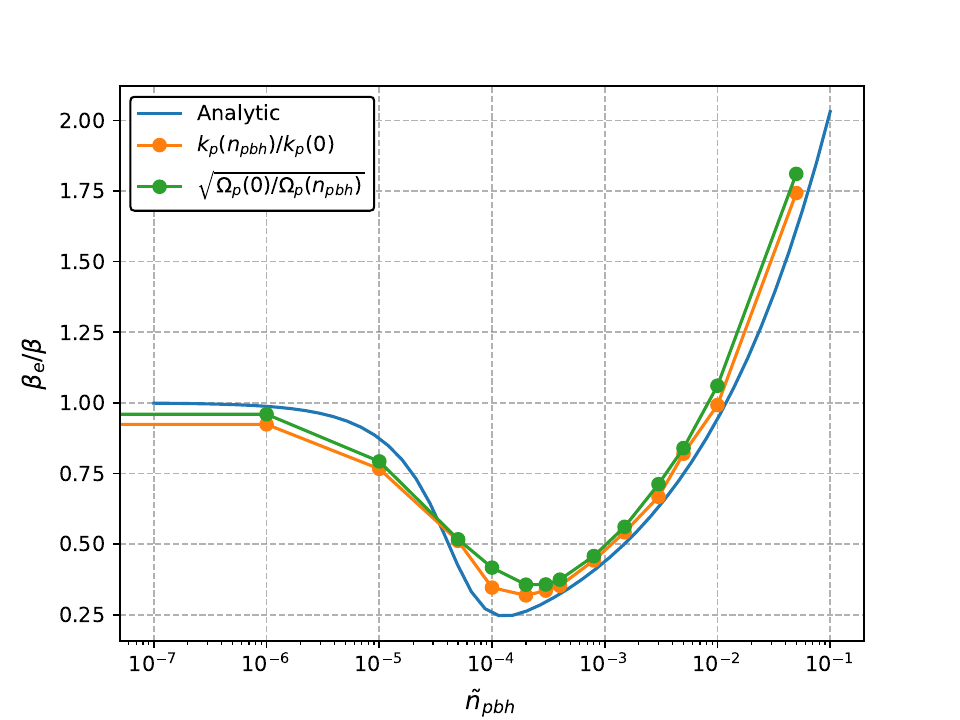}
\includegraphics[width=0.45\textwidth]{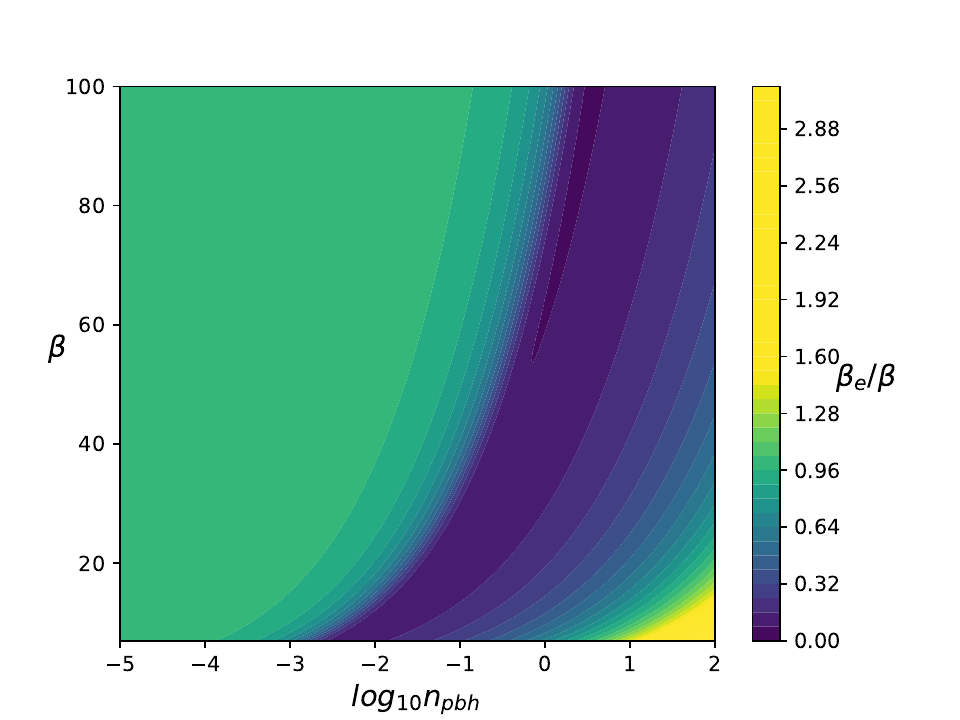}
\caption{ {\it Left:} Comparison between the numerical results and the analytical estimations from Eq.~\eqref{eq:betaeff} with \(G_0 H^4/\beta^4 = 10^{-8}\). X-coordinate label is \(\tilde{n}_{\rm pbh} = n_{\rm pbh} H^3/\beta^3\). We have chosen \(F(t_p) = 0.7\). {\it Right:} Analytical estimations of \(\beta_e/\beta\) in plane (\(n_{\rm pbh}\), \(\beta\)).  We have chosen \(G_0 = 10^{-10}\).}
\label{fig:betafit}
\label{fig:betaeff_diff_n}
\label{fig:betaeff_n_b}
\end{figure}

To go beyond these arguments on overall factors, we will discuss the spectral shape of GWs from different sources. Since we are interested in strong PTs, we will focus on bubble wall collision GWs and SIGWs.
}

\subsection{Bubble Collision GWs}
\label{subsec: PGW}
The spectral shape of bubble collision GWs depends on particular bubble collision profiles. In this study, we adopt two commonly used models——the analytical model with envelope approximation~\cite{jinno_gravitational_2017a,jinno_gravitational_2019} and the bulk flow model~\cite{Lewicki:2020azd,Lewicki:2022pdb}. Since the first model is analytically solvable, we will use the resultant GW spectrum from the first model to check whether the estimation Eq.~\eqref{eq:betaeff} works. For the bulk flow model, we will directly apply our estimation Eq.~\eqref{eq:betaeff}.

{\bf The analytical model.} In this model, thin-wall approximation and envelope approximation~\cite{Kosowsky:1991ua,Kosowsky:1992rz,Kosowsky:1992vn,Kamionkowski:1993fg} are assumed, which means that the thickness of the bubble wall can be neglected and the energy of the bubble wall would be released immediately after bubble wall collisions. In normal case without PBHs, the GW spectrum can be expressed as broken power-law 
\begin{align}
\Delta_{\rm collision}(k/k_p) = \Delta_p \frac{(a+b)^c}{\l(b\l[\frac{k}{k_p}\r]^{-\frac{a}{c}}+a\l[\frac{k}{k_p}\r]^{\frac{b}{c}}\r)^c}~,
\end{align}
where \(a=3,b=1,c=1.91\) are shape parameters, \(\Delta_p =0.039\) is the spectral peak value and \(k_p/\beta = 1.426\) is the peak wavenumber. Here the wavenumber \(k\) is defined as
\begin{align}
    \frac{k}{\beta} = \frac{1}{H}\l(\frac{a}{a_0}\r)^{-1} \l(\frac{H}{\beta}\r) 2\pi f=\frac{2\pi f}{1.65\times10^{-2}\ {\rm Hz}} \l(\frac{H}{\beta}\r) \l(\frac{0.1 \ {\rm GeV}}{T_{\rm re}}\r) \l(\frac{g_*}{100}\r)^{-\frac{1}{6}}~,
    \label{eq:k}
\end{align}
where \(f\) is the current GW frequency and \(a/a_0\) is the redshifted factor.

With the existence of PBHs, we can apply our estimation Eq.~\eqref{eq:betaeff} and thus 
\begin{align}
    \Omega_{\rm GW}(k) h^2= 1.67\times10^{-5}\l(\frac{g_*}{100}\r)^{-\frac{1}{3}} \l(\frac{H}{\beta_e}\r)^2  \Delta_{\rm collision}(k/\beta_e)~.
    \label{eq:bcGWs}
\end{align}
However, we can also directly calculate the GW spectrum if we extend the methods in Ref.~\cite{jinno_gravitational_2017a,jinno_gravitational_2019}. Under thin-wall and envelope approximation, GW spectrum can be expressed as the sum of single- and double-bubble contributions,
\begin{align}
    \Omega_{\rm GW}(k) h^2= 1.67\times10^{-5}\l(\frac{g_*}{100}\r)^{-\frac{1}{3}} \l(\frac{H}{\beta}\r)^2  \sum_i\Delta^{(i)}(k/\beta,G_0,n_{\rm pbh})~,
    \label{eq:bcGWs}
\end{align}
where \(i = s, d\) denote single- and double-bubble contribution, respectively. Here, to compare with the results in Ref.~\cite{jinno_gravitational_2017a}, we fix pre-fator as \((H/\beta)^2\) and thus \(\Delta^{(i)}\) are GW spectra depending on normalized PBH number densities and catalytic strength. The detailed derivations of the single- and double-bubble contributions are shown in App.~\ref{app:frame}. Final expression of single-bubble spectrum is 
\begin{align}
\Delta^{(s)} = \frac{2 \Tilde{k}^3}{ 3}\int_0^\infty \ddd \Tilde{r}\,
\int_0^{\Tilde{r}} \ddd \Tilde{t}_d\,
\int_{\Tilde{r}/2}^\infty \ddd \Tilde{T} \cos(\Tilde{k}\Tilde{t_d})  P(\tilde{T},\tilde{t}_d,\tilde{r})
\times 
\left[ j_0(\Tilde{k}\Tilde{r})K_0 + \frac{j_1(\Tilde{k}\Tilde{r})}{\Tilde{k}\Tilde{r}}K_1 + \frac{j_2(\Tilde{k}\Tilde{r})}{\Tilde{k}^2\Tilde{r}^2}K_2 \right]~,
\label{eq:D_single}
\end{align}
and double-bubble spectrum is 
\begin{align}
\Delta^{(d)} = &\frac{8\pi \Tilde{k}^3}{3} \int_0^\infty \ddd \Tilde{r}\,\int_0^{\Tilde{r}} \ddd \Tilde{t}_d\,
\int_{\Tilde{r}/2}^\infty \ddd \Tilde{T} \cos(\Tilde{k}\Tilde{t_d})  P(\tilde{T},\tilde{t}_d,\tilde{r})
\nonumber \\
&\times \left[G_0\frac{H^4}{\beta^4}C_0(\Tilde{r},\Tilde{T},\Tilde{t_d})+n_\mathrm{pbh}\frac{H^3}{\beta^3}C_1(\Tilde{r},\Tilde{T},\Tilde{t_d})\right] \times \left[G_0\frac{H^4}{\beta^4}C_0(\Tilde{r},\Tilde{T},-\Tilde{t_d})+n_\mathrm{pbh}\frac{H^3}{\beta^3}C_1(\Tilde{r},\Tilde{T},-\Tilde{t_d})\right].
\label{eq:D_double}
\end{align}
Here \(j_n\) are the spherical Bessel functions. \(K_n\) functions, \(P\) function and \(C_n\) functions are defined in App.~\ref{app:frame} which incorporate the influence from PBHs. Variables in formulas have been rescaled by \(\beta\),
\begin{align}
    \Tilde{k} = k/\beta~, \, \Tilde{r} = \beta r ~,\, \Tilde{t}_d = \beta t_d~,\, \Tilde{T} = T\beta~.
\end{align}


\begin{figure}[!tb]
\centering
\includegraphics[width=0.45\textwidth]{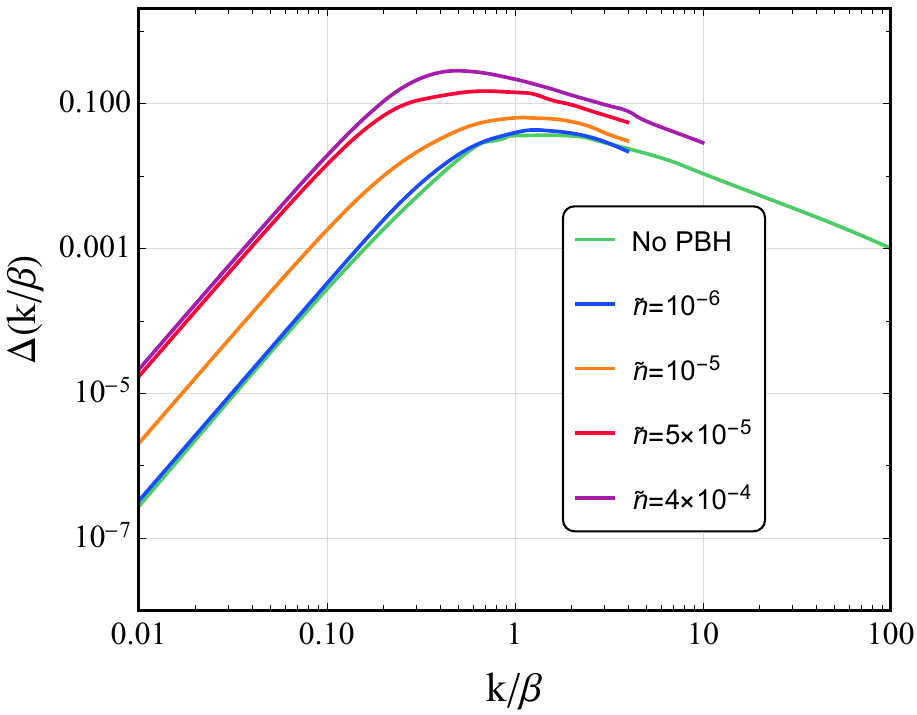}
\includegraphics[width=0.45\textwidth]{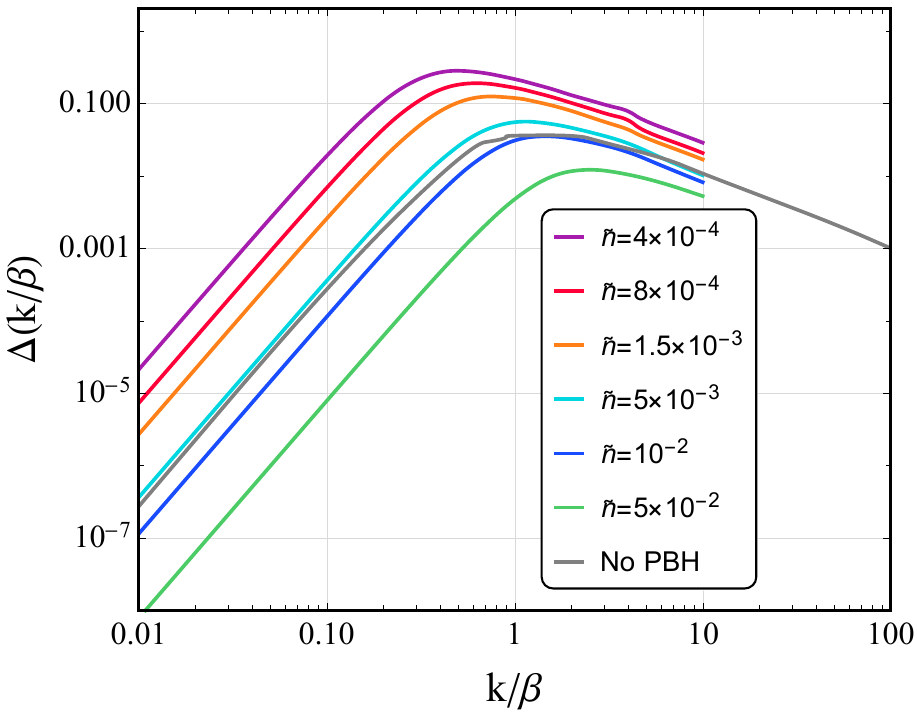}
\caption{GW spectra in various normalized PBH number densities.  Here \(\tilde{n}=n_{\rm pbh}H^3/\beta^3\) and we have chosen \(G_0H^4/\beta^4 =10^{-8}\).}
\label{fig:GWs}
\end{figure}

Numerical results of bubble collision GWs in the analytical model are shown in Fig.~\ref{fig:GWs}. We found that GW spectra show causal tails \(\propto k^3\) in the IR regime and \(\propto k^{-1}\) in the UV regime, which coincide with the standard case~\cite{jinno_gravitational_2017a}. Note that when normalized PBH number densities are sufficiently low, the PT dynamics are governed by background tunneling processes, reverting to the standard scenario~\cite{jinno_gravitational_2017a}. In addition, as already shown in the Left panel of Fig.~\ref{fig:betafit}, the numerical results coincide with the estimation Eq.~\eqref{eq:betaeff}. In the case \(G_0H^4/\beta^4 =10^{-8}\), GW spectra reach their maximum when \(n_{\rm pbh} H^3/\beta^3\sim  10^{-4}\), being approximately 16 times greater than the scenario without PBHs. Meanwhile, when \(n_{\rm pbh}H^3/\beta^3 \geq 10^{-2}\), the GW magnitudes become lower than the case without PBHs.

{\bf The bulk flow model.} This model includes the contributions from thin relativistic shells. We apply our estimation Eq.~\eqref{eq:betaeff} by substituting \(\beta\to\beta_e\) in the following formulas
\begin{equation}
\Omega_{\text{GW}}h^2 = {\cal N}\left( \frac{H}{\beta} \right)^{2} \frac{A(a + b)^c\, S_H(k, k_{H})}{\left[ b\,\left(\frac{k}{k_p}\right)^{-\frac{a}{c}} + a\,\left(\frac{k}{k_p}\right)^{\frac{b}{c}} \right]^c}~,
\label{eq:bfGWs}
\end{equation}
where \(k_p \approx 0.77 \beta/H\), \(A = 5.1\times 10^{-2},\ a=b=2.4,\ c=4\)~\cite{Lewicki:2022pdb}, and
\begin{equation}
S_H(k, k_{H}) = \left[ 1 + \frac{\Omega_{\text{CT}}(k_{H})}{\Omega_{\text{CT}}(k)} \left( \frac{k}{k_{H}} \right)^a \right]^{-1}~.
\end{equation}
Here the wavenumber \(k\) is defined as Eq.~\eqref{eq:k}. The function \(\Omega_{\rm CT}\) accounts for the causality-limited tail of the spectrum at \(k < k_H\). In pure radiation dominance, \(\Omega_{\rm CT} \propto k^3\)~\cite{Caprini_2009_general,Cai:2019cdl}. To get the current spectrum of GWs, the prefactor \({\cal N}\) needs to be \(1.67\times10^{-5}\ \l({g_*}/{100}\r)^{-\frac{1}{3}}\).

\subsection{Scalar-induced GWs}
\label{subsec: SIGW}
\begin{figure}[!tb]
\centering
\includegraphics[width=0.45\textwidth]{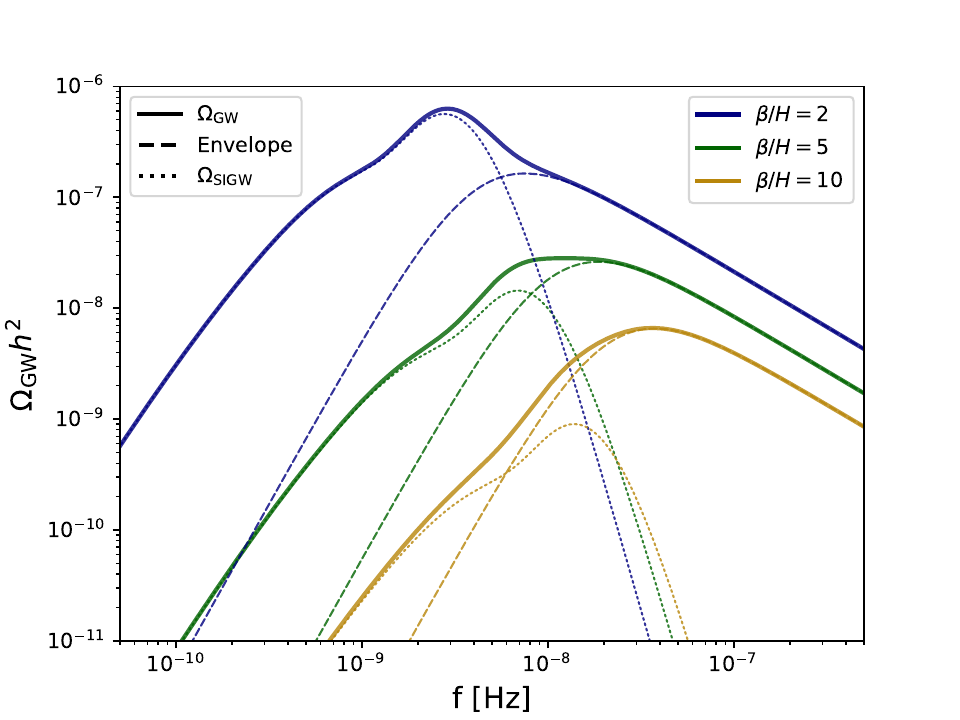}
\includegraphics[width=0.45\textwidth]{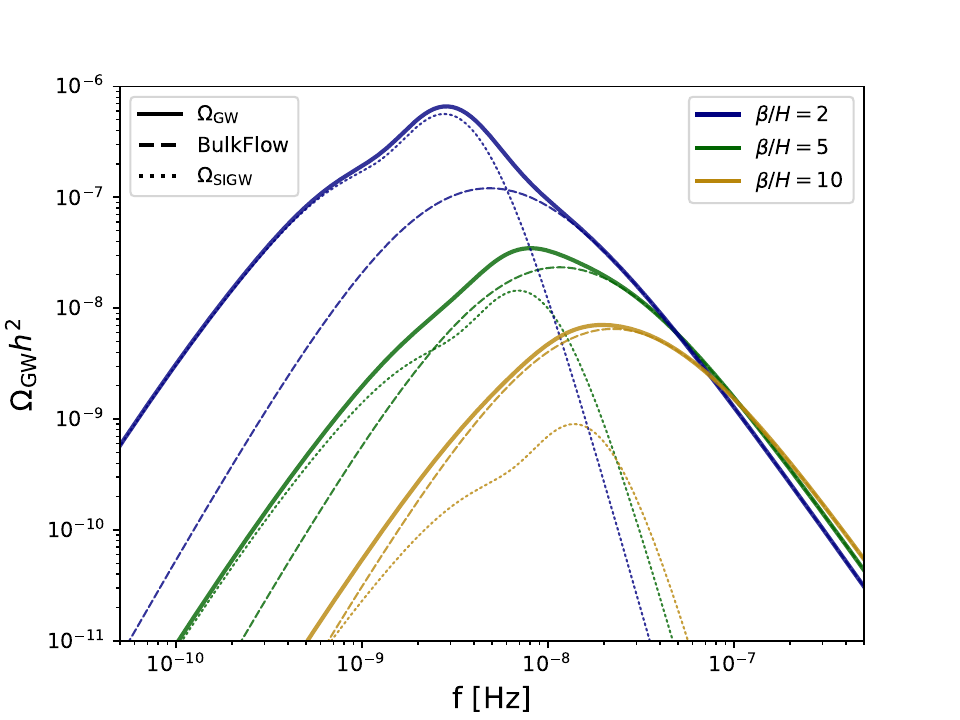}
\caption{The comparisons between the bubble collision GWs (dashed line) and SIGWs (dotted line) with various PT inverse durations \(\beta/H\). We have chosen \(T_{\rm re} = 0.1\ {\rm GeV}\).}
\label{fig:SIGW}
\end{figure}

The stochastic nature of bubble nucleation during a FOPT can also generate curvature perturbations, which will induce GW in the following evolution. We first simply review the PT accompanied SIGWs without PBHs' existence and then apply Eq.~\eqref{eq:betaeff} in a suitable way. The curvature perturbations power spectrum has been computed analytically in~\cite{jinno_curvature_2024} and numerically in~\cite{lewicki_black_2024,lewicki_black_2024a,franciolini_curvature_2025,zou_numerical_2025}.
Here we use the analytical results with Gaussian approximation from~\cite{jinno_curvature_2024}. The curvature perturbations originate from the asynchrony in the completion of PT in local patches,
\begin{equation}
\mathcal{P}_{\zeta} = \l(\frac{\alpha}{1+\alpha}\r)^2\mathcal{P}_{ \delta t}~,
\end{equation}
where \(\zeta\) is the curvature perturbation on uniform-density hypersurfaces and \({\cal P}_{\delta t}\) is the variation of local completion times of the PT in different patches. Since the process of PTs can convert vacuum energy into radiation energy, the variation of local completion times can provide isocurvature perturbations between vacuum and radiation energy density during the PT, which will evolute into curvature perturbations when the PT is completely finished~\cite{jinno_curvature_2024}. Here \(\alpha = \frac{\rho_V}{\rho_{rad}}\big|_{T_n} \gg 1\) in very strong PTs. The variation of local completion times can be expressed as
\begin{equation}
\mathcal{P}_{\delta t}(k) = \l(\frac{H}{\beta}\r)^2 \mathcal{P}_{\beta \delta t}(k/\beta)~,  
\end{equation}
where \( \mathcal{P}_{\beta \delta t}\) is the dimensionless power spectrum with \(\mathcal{P}_{\beta \delta t}(k/\beta) \approx 70\ (k/\beta)^3\) in the IR regime and \(\mathcal{P}_{\beta \delta t}(k/\beta)\approx 0.7\ (k/\beta)^{-3}\) in the UV regime. This spectrum can also be parameterized by a broken power-law~\cite{jinno_curvature_2024}
\begin{equation}
    \mathcal{P}_{\beta \delta t} = \Delta_p \frac{(a+b)^c}{\l(b\l[\frac{k}{k_p}\r]^{-\frac{a}{c}}+a\l[\frac{k}{k_p}\r]^{\frac{b}{c}}\r)^c}~,
\end{equation}
where \(a = b=3,\, c = 2.696\), \(k_p \approx 0.464\beta\) and \(\Delta_p = 1.08\).
The spectrum of the SIGWs reads~\cite{SIGW1,SIGW2}
\begin{align}
    \Omega_{\text{SIGW}}(k)h^2
= \mathcal{N} \int_{0}^{1} \mathrm{d}u \int_{1}^{\infty} \mathrm{d}s\,
{\cal T}_{\text{rad}}(u,s)\,
\mathcal{P}_\zeta\!\left(\frac{k}{2}(s+u)\right)\,
\mathcal{P}_\zeta\!\left(\frac{k}{2}(s-u)\right)~,
\end{align}
with~\cite{transfer1,transfer2}
\begin{align}
    {\cal T}_{\text{rad}}(u,s)
= 12\; \frac{(u^2-1)^2\, (s^2-1)^2\, (u^2+s^2-6)^4}{(s^2-u^2)^8}\,
\left[
    \left(
      \ln\!\left| \frac{3-u^2}{s^2-3} \right|
      + \frac{2(s^2-u^2)}{u^2+s^2-6}
    \right)^2
    + \pi^2\,\Theta(s-\sqrt{3})
\right]~.
\end{align}
To obtain the current GW spectrum, the prefactor \({\cal N}\) needs to be \(1.67\times10^{-5}\ \l({g_*}/{100}\r)^{-\frac{1}{3}}\). Here, we use package SIGWfast~\cite{Witkowski:2022mtg} to numerically evaluate SIGWs. Figure~\ref{fig:SIGW} shows the resulting GW spectra for various \(\beta/H\) with \(T_{\rm re} = 0.1\ {\rm GeV}\). We demonstrate that the peak frequency of SIGWs is lower than bubble collision GWs in both the analytical model with envelope approximation and the bulk flow model. Note that when \(\beta/H \lesssim 5\), SIGWs dominate over both the analytical model with envelope approximation and the bulk flow model. This is because SIGWs are more sensitive to the PT inverse duration, as \(\Omega_{\rm SIGW}\propto \l(H/\beta\r)^{4}\). Therefore, when PTs become slower (i.e., \(\beta/H\) is smaller), SIGWs experience greater enhancement compared to bubble collision GWs.

Influenced by PBHs, we expect the variation of local completion times to be modified in a manner analogous to bubble collision GWs, expressed as \(\mathcal{P}_{ \delta t}\to \mathcal{P}'_{ \delta t}(n_{\rm pbh}, G_0, \beta, k )\), with
\begin{align}
    {\cal P}'_{ \delta t}(n_{\rm pbh}, G_0, \beta, k) = \l(\frac{H_*}{\beta_e(n_{\rm pbh}, G_0,\beta)}\r)^2 {\cal P}_{\beta \delta t}(k/\beta_e)~.
\end{align}
As a consequence, SIGWs will be enhanced in the regime \(\beta_e/\beta <1\) and get suppressed when \(\beta_e/\beta>1\), which is similar to bubble collision GWs. The SIGW spectrum is
\begin{align}
        \Omega_{\text{SIGW}}h^2
= 1.67\times10^{-5}\l(\frac{g_*}{100}\r)^{-\frac{1}{3}} \l(\frac{H}{\beta_e}\r)^4\int_{0}^{1} \mathrm{d}u \int_{1}^{\infty} \mathrm{d}s\,
{\cal T}_{\text{rad}}(u,s)\,
\mathcal{P}_{\beta \delta t}\!\left(\frac{k(s+u)}{2\beta_e}\right)\,
\mathcal{P}_{\beta \delta t}\!\left(\frac{k(s-u)}{2\beta_e}\right)~.
\end{align}

\section{Results from PTA data}
\label{sec:influence}

In the previous analysis, we have quantified the catalytic effects of PBHs on PT GWs. Next, we will perform data fitting with the NANOGrav 15-year dataset~\cite{NANOGrav:2023gor,NANOGrav:2023hde,NANOGrav:2023hvm,nano_grav_2023_dataset}. For our data analysis, we employ the PT parameters \(T_{\rm re}\) and \(\beta/H\), which can be derived from the PT models through Eq.~\eqref{eq:PTmodel}, along with the PBH density parameter \(n_{\rm pbh}(0.1\ {\rm GeV})\), where \(0.1\ {\rm GeV}\) corresponds to the typical temperature of PT interpretation of PTA data~\cite{NANOGrav:2023hvm,ellis_what_2023,Gouttenoire_2023}. From Eq.~\eqref{eq:pbhdensity}, normalized PBH number density can be expressed by parameter \(T_{\rm re}\) and \(n_{\rm pbh}(0.1\ {\rm GeV})\) 
\begin{align}
n_{\rm pbh}(T_{\rm re}) = n_{\rm pbh}(0.1\ {\rm GeV})\l(\dfrac{0.1\ {\rm GeV}}{T_{\rm re}}\r)^{3}~.
\label{eq:para}
\end{align}
In the subsequent analysis, for brevity, we will denote \(n_{\rm pbh}(0.1\ \mathrm{GeV})\) simply as \(n_{\rm pbh}\). To ensure PT completion, we set \(\beta/H \geq 1\) in the subsequent analysis. We apply the Bayesian inference method to determine the best fit of bubble collision GW spectrum in Eq.~\eqref{eq:bcGWs} and Eq.~\eqref{eq:bfGWs} and of SIGW spectrum generated during PTs. We adopt the MCMCsampler {\bf emcee}~\cite{emcee} to sample the posterior probability. The priors of the reheating temperature \(T_{\rm re}\), PT inverse duration \(\beta/H\) and PBH density parameter \(n_{\rm pbh}\) follow log-uniform distributions within \(\log_{10}(T_{\rm re}/{\rm GeV})\in [-2,\ 2]\), \(\log_{10}(\beta/H)\in [0,\ 2]\) and \(\log_{10}(n_{\rm pbh})\in [-7,\ 10]\), respectively.

\begin{figure}[!tb]
\centering
\includegraphics[width=0.4\textwidth]{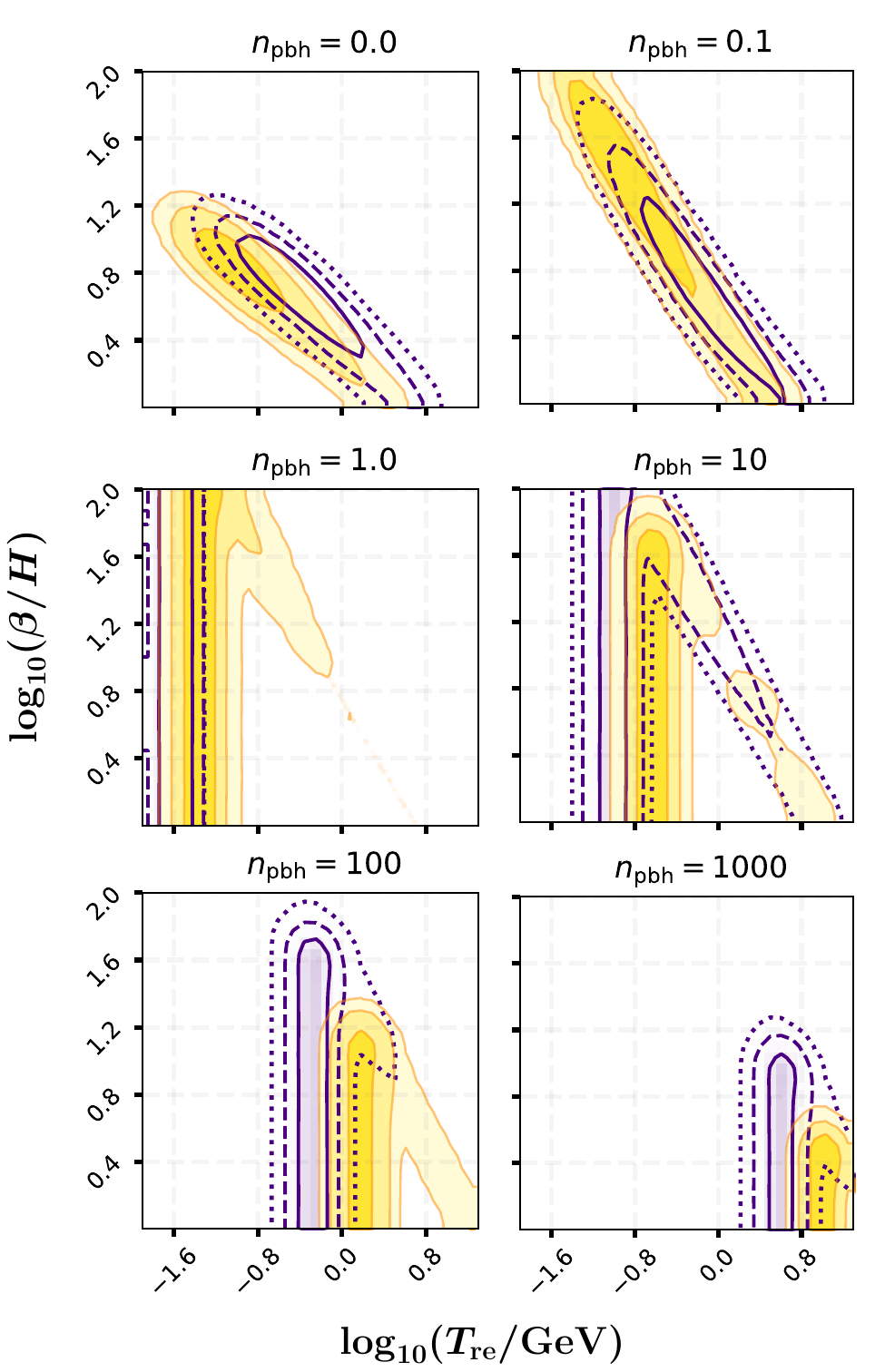}
\includegraphics[width=0.55\textwidth]{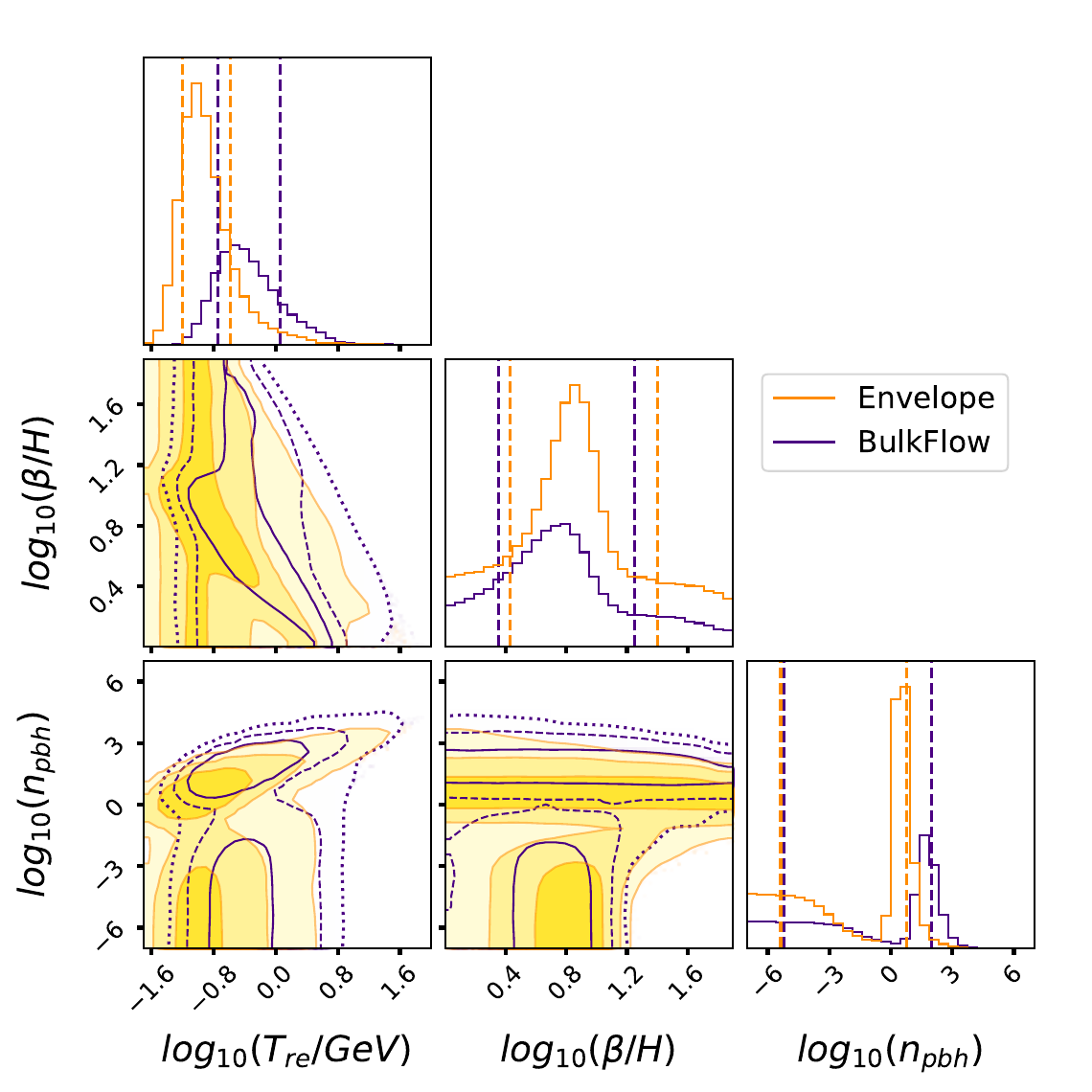}
\caption{The posterior probability distributions of bubble collision GWs in the analytical model with envelope approximation (yellow contours) and the bulk flow model (purple contours) fitting to the NANOGrav 15-year dataset~\cite{NANOGrav:2023hde,NANOGrav:2023gor,NANOGrav:2023hvm,nano_grav_2023_dataset}. We have chosen catalytic strength \(G_0=10^{-10}\) and denoted \(n_{\rm pbh}(0.1\ \mathrm{GeV})\) simply as \(n_{\rm pbh}\). For the analytical model with envelope approximation, the \(1\sigma,\, 2\sigma,\, \text{and } 3\sigma\) CL regions are depicted in progressively lighter shades of yellow, and for the bulk flow model, the same CL regions are enclosed by solid-, dashed- and dotted-lines. {\it Left:} The posterior probability distributions of PT parameters \(\beta/H\) and \(T_{\rm re}\) for given normalized PBH number densities \(n_{\rm pbh}\). {\it Right:} The posterior probability distributions of PT parameters \(\beta/H,\ T_{\rm re}\) and normalized PBH number densities \(n_{\rm pbh}\). On top of each column, we report \(1\sigma\) CL ranges.}
\label{fig:PTAPGW}
\end{figure}

\begin{figure}[!tb]
\centering
\includegraphics[width=0.4\textwidth]{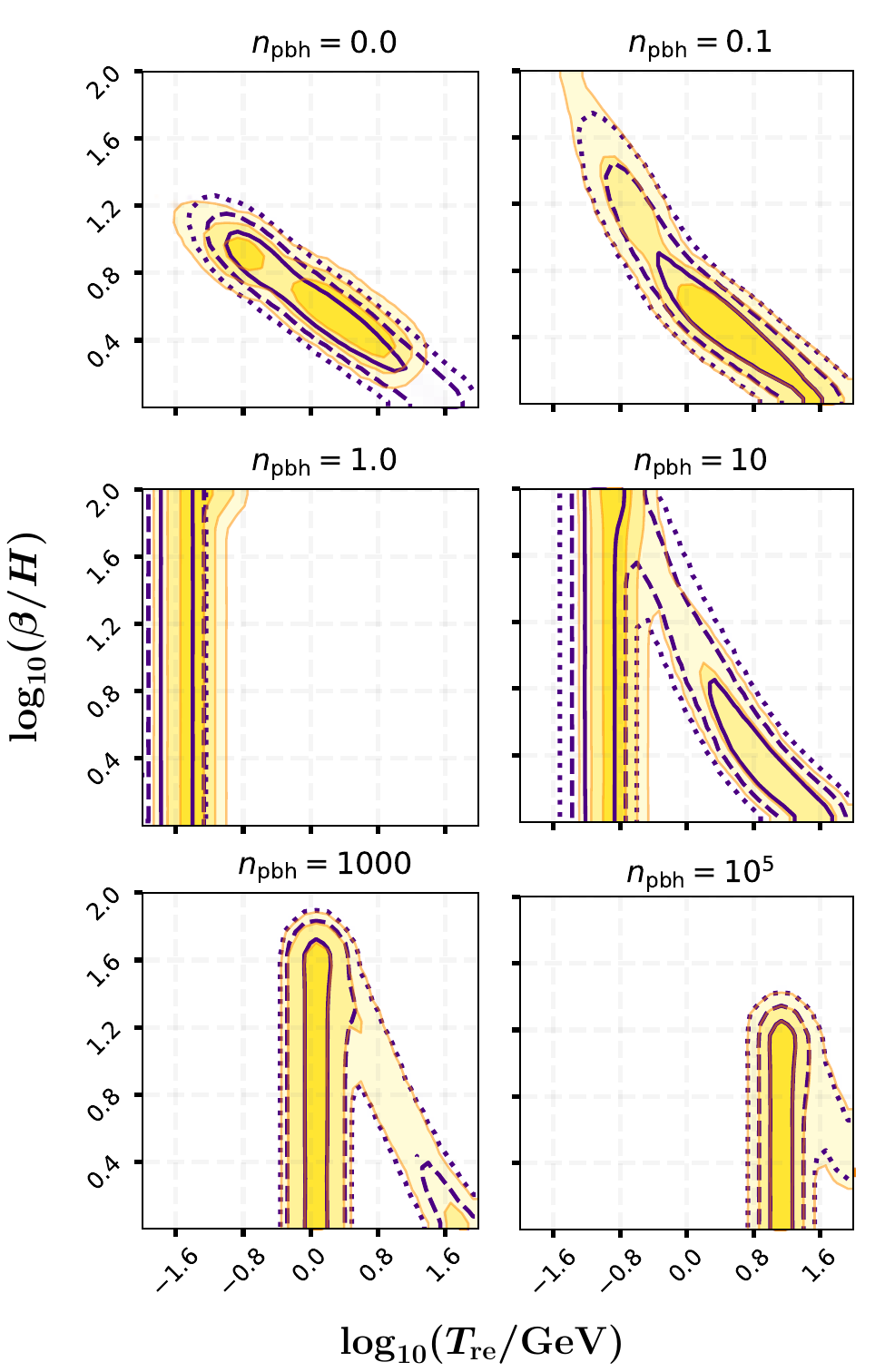}
\includegraphics[width=0.55\textwidth]{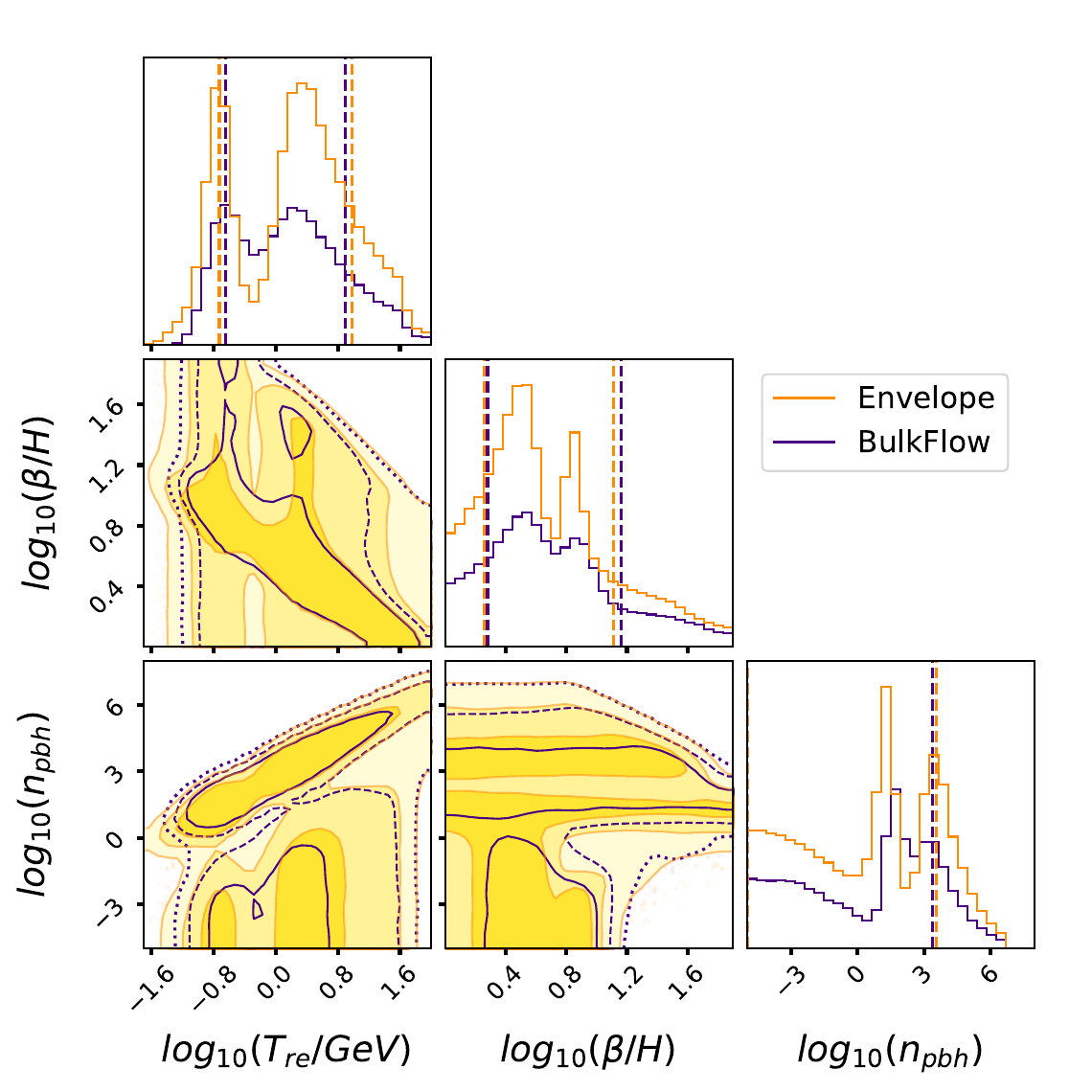}
\caption{The posterior probability distribution of combination with bubble collision GWs and SIGWs fit to the NANOGrav 15-year dataset~\cite{NANOGrav:2023hde,NANOGrav:2023gor,NANOGrav:2023hvm,nano_grav_2023_dataset} in Envelope+SIGW framework (yellow contours) and BulkFlow+SIGW framework (purple contours), respectively. We have chosen catalytic strength \(G_0=10^{-10}\) and denoted \(n_{\rm pbh}(0.1\ \mathrm{GeV})\) simply as \(n_{\rm pbh}\). For the analytical model with envelope approximation, the \(1\sigma,\, 2\sigma,\, \text{and } 3\sigma\) CL regions are depicted in progressively lighter shades of yellow, and for the bulk flow model, the same CL regions are enclosed by solid-, dashed- and dotted-lines. {\it Left:} The posterior probability distribution of PT parameters \(\beta/H\) and \(T_{\rm re}\) for given normalized PBH number densities \(n_{\rm pbh}\). {\it Right:} The posterior probability distribution of PT parameters \(\beta/H\), \(T_{\rm re}\) and normalized PBH number densities \(n_{\rm pbh}\). On top of each column, we report \(1\sigma\) CL ranges.}
\label{fig:PTAFGW}
\end{figure}

In Fig.~\ref{fig:PTAPGW}, we present a data fitting including only bubble collision GWs in the analytical model with envelope approximation and the bulk flow model, respectively. In the Left panel, for several given \(n_{\rm pbh}\), our analysis reveals significant PBH-induced modifications to PT parameter estimation.
At low normalized PBH number densities, the catalytic effect alters the correlation between \(T_{\rm re}\) and \(\beta\), manifesting as a parameter estimation bias toward higher \(\beta\) values. Interestingly, this aligns with previous observations where sparse PBH populations effectively reduce the equivalent inverse duration \(\beta\) (see Fig.~\ref{fig:betafit}). 
At high normalized PBH number densities, PBH-catalyzed PTs dominate the process, introducing substantial uncertainties in \(\beta\) determination. 
Note that the increasing normalized PBH number densities drive estimated parameters toward higher reheating temperatures. This arises because the normalized PBH number densities, which behave as \(n_{\rm pbh}(T_{\rm re}) \propto T_{\rm re}^{-3}\)
(see Eq.~\eqref{eq:para}), must remain suppressed to maintain a sufficiently slow PT so that the resulting GWs can be compatible with PTA-detected GW signals. Consequently, elevated reheating temperatures become necessary to preserve this density suppression. Meanwhile, adjustment of the PT inverse durations is also needed: higher \(T_{\rm re}\) increases the GW peak frequency \(f_p\), leading to the decrease of GW spectrum values at nanohertz since the causal IR tail scales as \(\Omega_{\rm GW}h^2 \propto ({\rm nHz}/f_p)^3\). Therefore an even slower PT (lower \(\beta/H\)) is required to amplify the GW spectrum for explaining the PTA signals. Such a slow PT is constrained by the fundamental PT completion requirement \(\beta/H \gtrsim 1\), ultimately imposing an upper limit on the permissible normalized PBH number densities.
The Right panel of Fig.~\ref{fig:PTAPGW} shows a joint estimation of PT parameters \(\beta,\ T_{\rm re}\) and normalized PBH number densities \(n_{\rm pbh}\). Our analysis reveals that the normalized PBH number densities are constrained below a critical threshold, with an upper limit at the \(3\sigma\) confidence level of \(n_{\rm pbh} \leq 10^{3.04}\) in the analytical model with envelope approximation and \(n_{\rm pbh} \leq 10^{3.73}\) in the bulk flow model.

In Fig.~\ref{fig:PTAFGW}, we perform the same data fitting using a combination with SIGWs and bubble collision GWs in the analytical model with envelope approximation (envelope+SIGW) and the bulk flow model (bulk flow+SIGW), respectively. In the Left panel of Fig.~\ref{fig:PTAFGW}, the same trend of parameter estimation in different normalized PBH number densities persists when incorporating the contributions from SIGWs.
In the absence of PBHs, there are two \(1\sigma\) regions in the envelope+SIGW framework. These regions correspond to the SIGW peak (higher \(T_{\rm re}\)) and the bubble collision GW peak (lower \(T_{\rm re}\)), respectively. However, in the bulk flow+SIGW framework, these two \(1\sigma\) regions cannot be distinguished because the SIGW peak is closer to the bubble collision GW peak in the bulk flow model (see Fig.~\ref{fig:SIGW}).
Note that when normalized PBH number densities are larger than \(n_{\rm pbh} \gtrsim 10^3\), parameter estimation results of both the envelope+SIGW framework and the bulk flow+SIGW framework are almost the same. This arises because SIGWs dominate the full GW spectrum at sufficiently slow PTs (\(\beta/H \lesssim 5 \), see Fig.~\ref{fig:SIGW}). From the joint estimation in the Right panel of Fig.~\ref{fig:PTAFGW}, the upper limit at the \(3\sigma\) confidence level of the normalized PBH number densities is relaxed to \(n_{\rm pbh} \leq 10^{6.51}\) under the envelope+SIGW framework and \(n_{\rm pbh} \leq 10^{6.49}\) under the bulk flow+SIGW framework. 
This relaxation occurs because SIGWs provide additional flexibility in matching PTA data. As discussed in Sec.~\ref{subsec: SIGW}, the SIGW spectral peak occurs at lower frequencies compared to the bubble collision GW spectral peak, and the rising edge of the SIGW peak can also explain PTA data if it dominates the full GW spectrum. Explanation of PTA signals with PT SIGWs corresponds to higher PT reheating temperatures compared to explanations involving only bubble collision GWs. Therefore, the normalized PBH number densities are suppressed as \(n_{\rm pbh}(T_{\rm re})\propto T_{\rm re}^{-3}\), allowing for larger \(n_{\rm pbh}(0.1\ {\rm GeV})\).

\begin{figure}[!tb]
\centering
\includegraphics[width=0.45\textwidth]{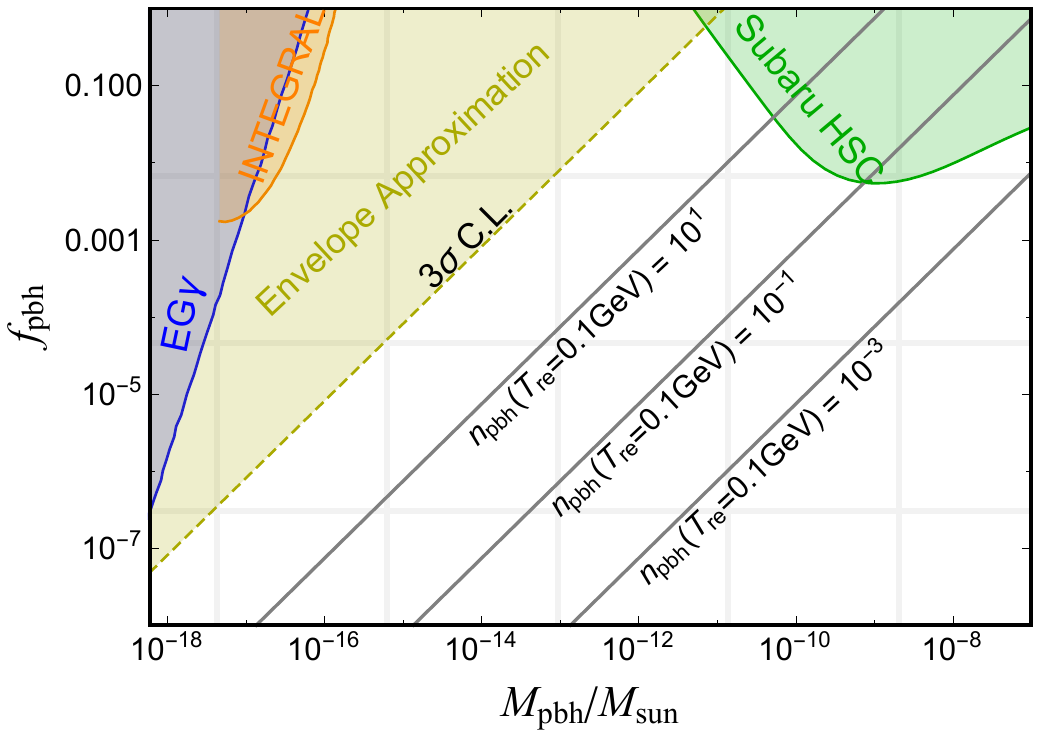}
\includegraphics[width=0.45\textwidth]{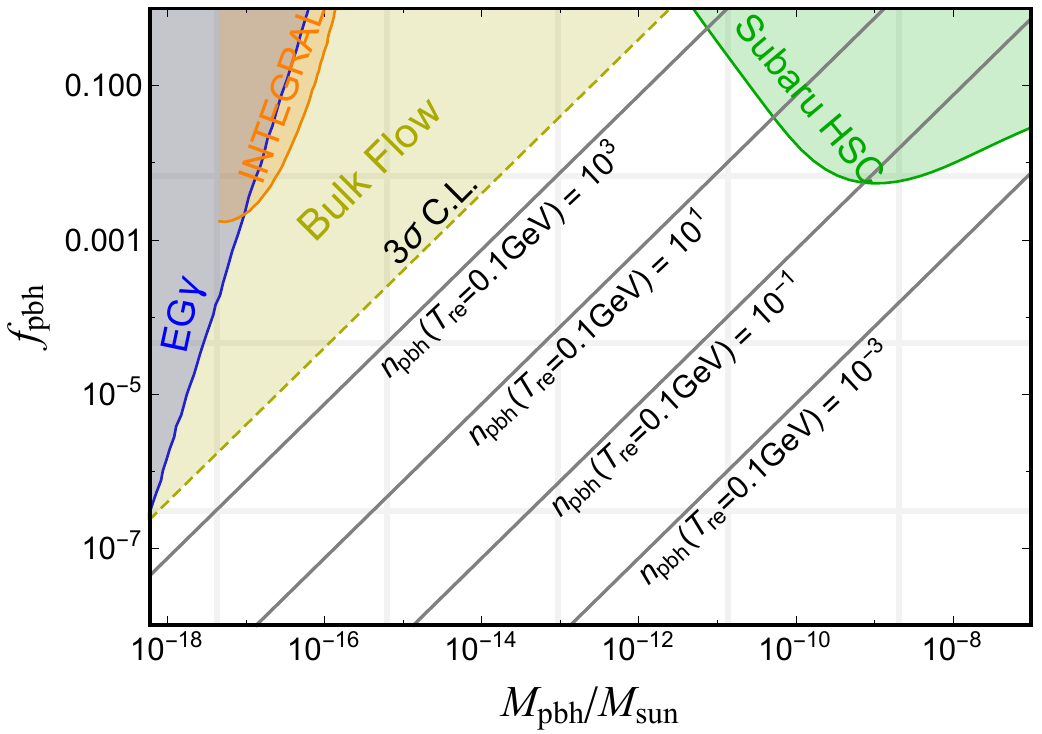}
\includegraphics[width=0.45\textwidth]{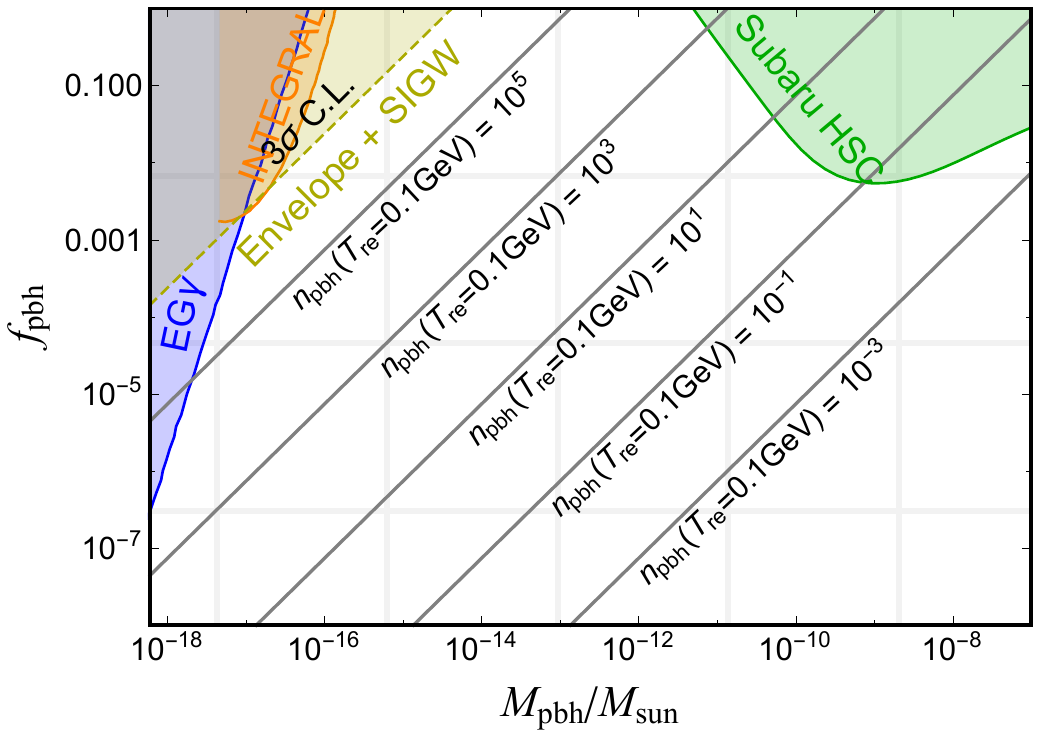}
\includegraphics[width=0.45\textwidth]{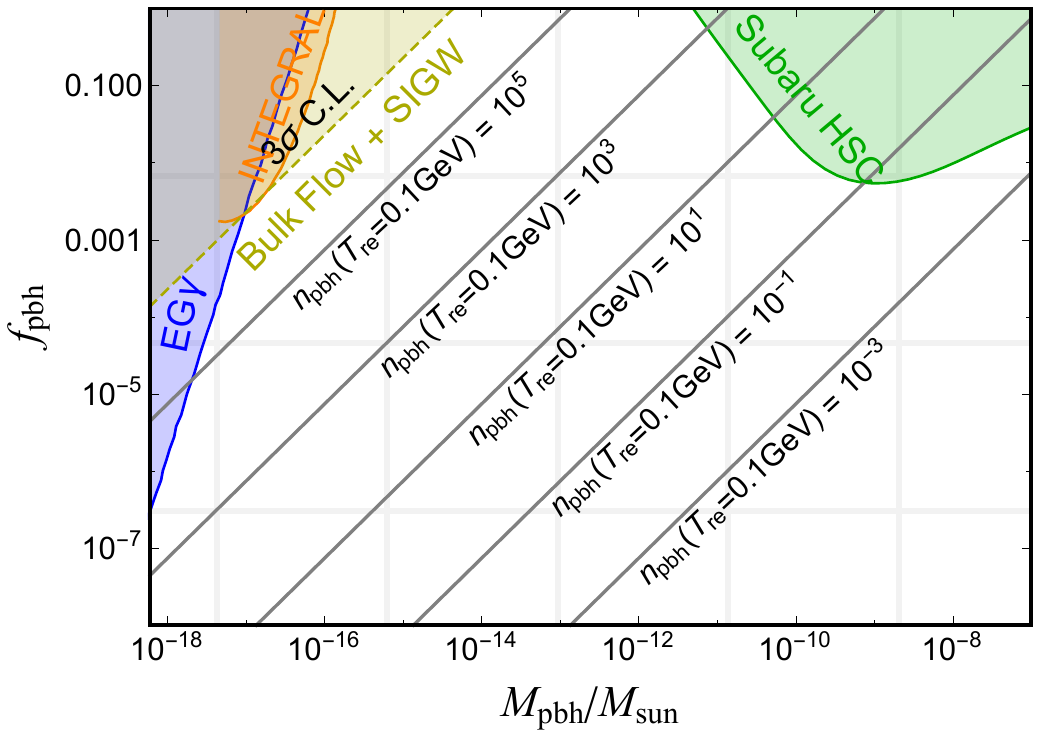}
\caption{Incompatibility of asteroid-mass PBHs with PT interpretation of PTA data. We present \(3\sigma\) CL constraint region in different models respectively. Gray lines represent constant value of different \(n_{\rm pbh}(0.1 {\rm GeV})\) in PBH parameter space. In addition, we present the constraints from the SUBARUHSC~\cite{HSC1,HSC2}, the Hawking evaporation producing extra-galactic gamma-ray (EG \(\gamma\))~\cite{EG} and the gamma-ray observations by INTEGRAL (INT)~\cite{INT}.}
\label{fig:pbhconstraints}
\end{figure}


{
Previous analyses have illustrated how PBHs influence the PT interpretation of PTA. We now focus on the compatibility between the asteroid-mass PBH dark matter hypothesis and the PT interpretation of PTA data. It is notable that in each model, there is an upper limit on parameter \(n_{\rm pbh}(0.1\ {\rm GeV})\), implying that if the observed SGWB originates from a FOPT, a constraint on \(n_{\rm pbh}(0.1\ {\rm GeV})\) will be obtained. The parameter \(n_{\rm pbh}(0.1\ {\rm GeV})\) is a function of the mass of PBH and present PBH mass fraction (See Eq.~\eqref{eq:pbhdensity}). 
}
In Fig.~\ref{fig:pbhconstraints}, we present the \(3\sigma\) C.L. constraints of \(n_{\rm pbh}(0.1 \ {\rm GeV})\) in the PBH parameter space, evaluated through four distinct models: analytical model with envelope approximation, bulk flow model, and their combinations with SIGWs (envelope+SIGW and bulk flow+SIGW). Gray lines show the corresponding values of current PBH mass fraction \(f_{\rm pbh}\) as a function of PBH mass \(M_{\rm pbh}\) in several given \(n_{\rm pbh}(0.1\ {\rm GeV})\). Relations between these quantities can be seen in Eq.~\eqref{eq:pbhdensity}. Under both the analytical model with envelope approximation and the bulk flow model, we find that almost all possibilities of asteroid-mass PBHs acting as dark matter are excluded. After considering the impact of SIGWs from PTs, the constraint on \(10^{-14}-10^{-12}M_{\odot}\) PBH masses is relaxed, but \(10^{-16}-10^{-14}M_{\odot}\) PBHDM is still in conflict with the PT interpretation of PTA signals. 
{
The asteroid-mass PBHDM is also associated with SIGW if they were generated through large overdensities~\cite{sasaki_primordial_2018}. However, the SIGW from PBH production will not affect the result since the peak frequencies are much higher. See App.~\ref{app:sigw} for more details.
}

However, the precise contribution of SIGWs remains under debate~\cite{franciolini_curvature_2025,lewicki_black_2024,jinno_curvature_2024,zou_numerical_2025}. Notably, if we adopt the results from Ref.~\cite{franciolini_curvature_2025}, where SIGWs remain subdominant in strong PTs, considering SIGWs should not alter the results obtained by data fit with only bubble collision GWs. This implies that PBHDM in the mass range \(10^{-14}M_\odot-10^{-12}M_\odot\) would also conflict with the PT interpretation of PTA signals.
{
Compared to the methods in Ref.~\cite{franciolini_curvature_2025}, our result, based on methods in Ref.~\cite{jinno_curvature_2024}, yields a looser constraint on asteroid-mass PBHDM. 
A fully consistent PTA likelihood using the high-resolution simulations of Ref.~\cite{franciolini_curvature_2025} lies beyond the present scope and will be addressed in forthcoming work dedicated to the numerical modeling of SIGWs.
}

\section{Conclusion and Discussions}
\label{sec:disc}
The possibilities of primordial black hole dark matter (PBHDM) and the PT interpretation of PTA GW signals have been attracting significant attention in the recent literature~\cite{PBHdarkmatter,Bird:2016dcv,Clesse:2016vqa,Sasaki:2016jop,Carr:2020xqk,LISACosmologyWorkingGroup:2023njw,Gouttenoire:2023bqy,ellis_what_2023,NANOGrav:2023hvm,Gouttenoire:2025wxc,Salvio:2023blb,Salvio:2023ynn,Athron:2023mer}. Notably, the catalytic effect of PBHs on cosmological PTs has been widely acknowledged~\cite{Burda2015VacuumMetastabilitywithBlackHoles,Canko2018OntheCatalysisoftheElectroweakVacuumDecaybyBlackHolesatHighTemperature,Gregory2014BlackHolesAsBubbleNucleationSites,Gregory2016TheFateoftheHiggsVacuum,Gregory2020BlackHolesOscillatingInstantonsandtheHawkingMossTransition,Hiscock1987CanBlackHolesNucleateVacuumPhaseTransitions,Kohri2018ElectroweakVacuumCollapseInducedbyVacuumFluctuationsoftheHiggsFieldaroundEvaporatingBlackHoles,Moss1985BlackHoleBubbles,Mukaida2017FalseVacuumDecayCatalyzedbyBlackHoles,Strumia2023BlackHolesDon’tSourceFastHiggsVacuumDecay,oshita_compact_2019,jinno_superslow_2024,zeng_phase_2024}. 
Does PBHDM align with the PT-based explanation for PTA signals? We explore this intriguing question by investigating the catalytic effect of PBHs on GWs from very strong PTs.
PBHs can accelerate the PT process by nucleating earlier bubble formation around their vicinity, generating larger bubbles that expedite the PT. Using dimensional analysis and employing the analytical model with envelope approximation, we calculated GW signals from bubble collisions in the presence of PBHs. General expressions of bubble collision GWs are derived in Eq.~\eqref{eq:bcGWs}. Our results demonstrate that at relatively small normalized PBH number densities, the GW signals are enhanced due to the existence of large-size bubbles, while higher normalized PBH number densities suppress GW signals since the accelerated PT progresses too rapidly. We further extended to bubble collision GWs in the bulk flow model and SIGWs generated during PTs, establishing a unified framework for estimating GW signatures in PT scenarios. 

In addition, we used the NANOGrav 15-year dataset to analyze bubble collision GWs and their combination with SIGWs in the presence of PBHs. Our results reveal that the presence of PBHs introduces significant uncertainties in estimating PT parameters  (see Fig.~\ref{fig:PTAPGW} and Fig.~\ref{fig:PTAFGW}). Furthermore, if the PTA-detected SGWB is attributed to a cosmological PT, a substantial portion of the asteroid-mass PBH parameter space will be excluded (see Fig.~\ref{fig:pbhconstraints}). 
When considering only bubble collision GWs, nearly all possibilities of asteroid-mass PBHDM will be excluded.
This occurs because if all dark matter is composed of asteroid-mass PBHs, the averaged PBH number per Hubble volume at \(T\sim 0.1\ {\rm GeV}\) exceeds \(10^3\) (see Fig.~\ref{fig:pbhconstraints}), significantly accelerating the PT completion due to the PBH catalytic effect. This results in a weaker GW spectrum, with \(\Omega_{\rm GW}h^2 \ll 10^{-9}\).
Nevertheless, when incorporating the contributions from SIGWs, these constraints are relaxed, and only a portion of asteroid-mass PBHDM is excluded. However, our results rely on analytical results obtained with the Gaussian approximation~\cite{jinno_curvature_2024}. The impact of SIGWs on the resulting constraints depends on the specific model used. Further investigations are required to confirm how SIGWs from catalyzed PTs affect the constraints on asteroid-mass PBHDM.

In this study, we focused exclusively on bubble collision GWs and SIGWs, deliberately excluding sound wave and turbulence sources from our analysis. While this approach suffices for modeling very strong PTs and explaining the PTA-observed GW signals, we acknowledge that the sound wave and turbulence contributions become essential when investigating weaker PTs or preparing for future GW detectors with improved sensitivity. We leave the potential impact of PBHs on sound wave and turbulence-generated GWs for future investigations.
In this work, we only consider PBHs as catalysts, and it is straightforward to apply our formalism to other cases of impurity-catalyzed FOPTs.

\acknowledgments

We are grateful to Dongdong Zhang and Xin-Chen He for insightful comments.
This work was supported in part by the National Key
R\&D Program of China (2021YFC2203100), by the National Natural Science Foundation of China (12433002, 12261131497, 125B1023), by CAS young interdisciplinary innovation team (JCTD-2022-20), by 111 Project (B23042),
by CSC Innovation Talent Funds, by USTC Fellowship
for International Cooperation, and by USTC Research Funds of the Double First-Class Initiative. 
C.C. thanks the Peking University during his visit.

\appendix
\section{Bubble Collision GWs in the Analytical Model}
\label{app:frame}

Here we present the detailed derivations of GWs from bubble collisions in the framework\footnote{This framework also neglect cosmic expansion during PTs. See~\cite{Zhong:2021hgo} for modification from cosmic expansion.} of~\cite{jinno_gravitational_2017a,jinno_gravitational_2019}. The general expression of the GW spectrum, using the Green function method is

\begin{align}
	\Omega_{\rm GW}(t, k) &=\frac{1}{\rho_{tot}}\frac{\ddd \rho_{\rm GW}(t,k)}{\ddd \ln k} \notag
    \\
    &=\frac{2Gk^3}{ \pi\rho_{tot}} \int_{t_{\text{start}}}^{t_{\text{end}}} \ddd t_x 
\int_{t_{\text{start}}}^{t_{\text{end}}} \ddd t_y  \cos[k(t_x-t_y)] \Pi(t_x, t_y, k) ~,
\end{align}
where \(\Pi(t_x,\ t_y\ k)\) is the two-point correlation function of energy-momentum tensor
\begin{align}
     \Pi(t_x, t_y, k) = K_{ij,kl}(\hat{k}) K_{ij,mn}(\hat{k}) \int d^3 r\ e^{i\vec{k} \cdot \vec{r}} \langle T_{kl}(t_x, \vec{x}) T_{mn}(t_y, \vec{y}) \rangle~,
\end{align}
with \(\vec{r} \equiv \vec{x}-\vec{y} \), where the braket represents ensemble average of sources.
The current GW spectrum can be expressed as 
\begin{align}
    \Omega_{\rm GW}( k) h^2 = 1.67\times10^{-5}\ \l(\frac{g_*}{100}\r)^{-\frac{1}{3}} \Omega_{\rm GW}(t, k)~.
\end{align}

Assuming thin-wall approximation (the width of bubble wall can be neglected) and runaway profile (bubble wall propagate with speed of light, \(v_w\sim 1\)), energy-momentum tensor of the uncollided wall of a single bubble nucleated at $x_N = (t_N, \vec{x}_N)$ can be written as
\begin{equation}
	T_{ij}(x) = \rho(x) (\widehat{\vec{x} - \vec{x}_N})_{i} (\widehat{\vec{x} - \vec{x}_N})_{j}~,
\end{equation}
with
\begin{equation}
	\rho(x) = \l\{
	\begin{matrix}
		{4\pi \over 3} r_B(t)^3 {\kappa \rho_V \over 4\pi r_B(t)^2 l_B}~, 
		& r_B(t) < |\vec{x} - \vec{x}_N| < r_B(t) + l_B
		\\
		\\
		0~, & \text{otherwise}
	\end{matrix}
	\r.
\end{equation}
and
\begin{equation}
	r_B(t) = v_w(t - t_N) ~.
\end{equation}
Here, $l_B$ is the width of the wall, \(r_B\) is radius of bubble satisfied \(l_B \ll r_B\) and \(\rho_V\) is vacuum energy density. For very strong PTs, \(\rho_{\rm tot} = \rho_V\) and energy fraction \(\kappa=1\). Also, we assume that the energy-momentum tensor vanishes once they collide with others, namely the envelope approximation. Consequently, any spacial point is passed by bubble walls only once and necessary condition for non-zero two-point correlation function is that spacetime (\(x,y\)) both remain in the false vacuum before (\(t_x,\ t_y\)), respectively. The survival probability for two spacetime points \(P(t_x,t_y, r)\) can be calculated as
\begin{equation}
    P(t_x,t_y,r) = e^{-I(x,y)}~,\quad I(x,y)= \int_{V_{xy}}d^4z \Gamma(z)~,
    \label{eq:surv1}
\end{equation} 
where \(V_{xy}\) denotes the union of past line cone of two spacetime points \(V_{xy} = V_x \cup V_y\). Nucleation rate \(\Gamma(z)\) is given in Eq.~\eqref{eq:nucleationrate} with \(t_c=t_\star =  0\). In the following, we sometimes use variables (\(T,\, t_d,\,r\)) defined as
\begin{align}
    T = \frac{t_x+t_y}{2}~,\quad t_d = t_x-t_y~, \quad r = |\vec{x}-\vec{y}|~.
\end{align}
Also we define the dimensionless variables as 
\[\Tilde{k} = k/\beta\ , \, \Tilde{r} = \beta r\ ,\, \Tilde{t}_d = \beta t_d~,\, \Tilde{T} = T\beta\ .\]

Explicit form of \(I(x,y)\) can be derived as
\begin{align}
I(x,y) &= G_0\frac{H^4}{\beta^4}8\pi e^{\tilde{T}}\mathcal{I}(\Tilde{t}_d,\Tilde{r})+n_\mathrm{pbh}\frac{H^3}{\beta^3} I'(\Tilde{r},\Tilde{T},\Tilde{t}_d)~, \label{eq:surv2}
\\
{\cal I}(\tilde{t}_d,\tilde{r}) &= e^{\tilde{t}_d/2}+e^{-\tilde{t}_d/2}+\frac{\tilde{t}_d^2-(\tilde{r}^2+4\tilde{r})}{4\tilde{r}}e^{-\tilde{r}/2}~,
\\
I'(\Tilde{r},\Tilde{T},\Tilde{t}_d) &=
\left\{\begin{matrix} 
    -\dfrac{\pi (\Tilde{r}+2\Tilde{T})^2(\Tilde{r}^2-4\Tilde{r}\Tilde{T}-3\tilde{t}_d^2)}{12\Tilde{r}} \quad & \Tilde{T}>\Tilde{r}/2 \\[2ex]  
  \frac{2}{3}\pi \Tilde{T}(4\Tilde{T}^2+3\tilde{t}_d^2)\quad & \Tilde{T}<\Tilde{r}/2
\end{matrix}\right.~,
\end{align}
where the first term in Eq.~\eqref{eq:surv2} coincide with~\cite{jinno_gravitational_2017a} represent scenarios without PBHs. 

Bubbles nucleated at the lateral surface of the cone \(V_x\)(\(V_y\)) can pass through spacetime point \(x\)(\(y\)) in the future. We denote such regions as \(\delta V_x\)(\(\delta V_y\)) and their width is given by bubble wall width \(l_B\). 
The two-point correlation function can then take only two possible forms, named the single- and double-bubble contributions, respectively. The single-bubble contribution refers to scenarios where the energy-momentum tensor at both locations originates from a single bubble wall, whereas the double-bubble contribution describes situations where these two spacetime points receive energy-momentum tensor components from two distinct bubble walls associated with separate bubbles. See Ref.~\cite{jinno_gravitational_2017a} for detailed explanations. Therefore the GW spectrum can be decomposed into
\begin{align}
    \Omega_{\rm GW}(t, k)= \l(\frac{H}{\beta}\r)^2  \sum_i\Delta^{(i)}(k/\beta, G_0, n_{\rm pbh})~,
\end{align}
where \(i = s, d\) denote single- and double-bubble contribution, respectively. Spectra \(\Delta^{(i)}\) can be expressed as
\begin{align}
    \Delta^{(i)} = \frac{3}{4\pi^2} \frac{\beta^2 k^3}{ \rho_{V}^2} 
\int_{t_{\text{start}}}^{t_{\text{end}}} \ddd t_x 
\int_{t_{\text{start}}}^{t_{\text{end}}} \ddd t_y 
\cos\big(k(t_x - t_y)\big) \Pi^{(i)}(t_x, t_y, k)~.
\end{align}


\subsection{Single-Bubble Contribution}

Two conditions are required for one single bubble contributes nonvanishing two-point correlation function: (i) No bubble nucleates inside \(V_{xy}\) ; (ii) One bubble nucleates in \(\delta V_{xy}\) , where \(\delta V_{xy}\) is cross-section of the lateral surfaces of past line cone of two distinct spacetime points \(\delta V_{xy} = \delta V_x \cap \delta V_y\) . Ensemble average gives
\begin{align}
\langle T_{kl}(t_x, \vec{x}) T_{mn}(t_y, \vec{y}) \rangle^{(s)}
&= 
P(t_x,t_y,r) \int_{-\infty}^{t_{xy}} dt_n \Gamma(t_n) 
{\mathcal T}^{(s)}_{ij,kl}(t,t_x,t_y,\vec{r})~,
\label{eq:Single_TT}
\end{align} 

where ${\mathcal T}^{(s)}_{ij,kl}$ is the value of $T_{ij}(x)T_{kl}(y)$ generated
by the wall of the bubble nucleated at time $t_n$.
This is calculated as
\begin{align}
{\mathcal T}^{(s)}_{ij,kl}
&= 
\left( \frac{4\pi}{3} r_x(t_n)^3 \cdot \rho_0 \cdot \frac{1}{4\pi r_x(t_n)^2l_B} \right) 
\left( \frac{4\pi}{3} r_y(t_n)^3 \cdot \rho_0 \cdot \frac{1}{4\pi r_y(t_n)^2l_B} \right)\int_{R_{xy}} d^3z \; (N_{\times}(t_n))_{ijkl}~,
\end{align}
with $(N_{\times})_{ijkl} \equiv (n_{x\times})_i (n_{x\times})_j (n_{y\times})_k (n_{y\times})_l$, where \(n_{x\times}(n_{y\times})\) is a unit vector from nucleation site to point \(x(y)\).
Here $R_{xy} \equiv \delta V_{xy} \cap \Sigma_{t_n}$ is the intersection of \(\delta V_{xy}\) and constant-time hypersurface \(\Sigma_{t_n}\) at time \(t
_n\). Since there are no special directions except for $\vec{r}$, $\langle T_{kl}(t_x, \vec{x}) T_{mn}(t_y, \vec{y}) \rangle^{(s)}$  can be decomposed as follows:
\begin{align}
   \langle T_{kl}(t_x, \vec{x}) T_{mn}(t_y, \vec{y}) \rangle^{(s)} &= a_1 \delta_{ij} \delta_{kl}
+ a_2 \frac{1}{2}(\delta_{ik}\delta_{jl} + \delta_{il}\delta_{jk}) 
+ b_1\delta_{ij}\hat{r}_k\hat{r}_l
+ b_2\delta_{kl}\hat{r}_i\hat{r}_j \nonumber \\
&\;\;\;\;+ b_3 \frac{1}{4}(\delta_{ik}\hat{r}_j\hat{r}_l + \delta_{il}\hat{r}_j\hat{r}_k
+ \delta_{jk}\hat{r}_i\hat{r}_l + \delta_{jl}\hat{r}_i\hat{r}_k) 
+c_1 \hat{r}_i\hat{r}_j\hat{r}_k\hat{r}_l~,
\end{align}

where \(a_i,b_i,c_i\) are scalar functions over (\(t_d, T,r\)). After the projection by $K$, only a few terms survive:
\begin{align}
&K_{ij,kl}(\hat{k})K_{ij,mn}(\hat{k})
\langle T_{kl}T_{mn} \rangle^{(s)} 
= 2a_2 
+ \l(1-\l(\hat{r} \cdot \hat{k}\r)^2\r)b_3 
+ \frac{1}{2}\l(1-\l(\hat{r} \cdot \hat{k}\r)^2\r)^2c_1~,
\end{align}
We neglect the contribution from \(t_{n}<0\) in the subsequent analysis since we have assumed \(G_0 \ll 1\) in Eq.~\eqref{eq:nucleationrate}. This, however, requires that \(t_{xy}>0\), i.e., \(T>\dfrac{r}{2}\). After integration, we can calculate \(a_2,b_3,c_1\) as
\begin{align}
    a_2 &= \frac{\pi}{18}\rho_V^2 P(t_x,t_y,r) K_0/r^2~,
    \\
    b_3 &= \frac{\pi}{36}\rho_V^2 P(t_x,t_y,r) K_1/r^2~,
    \\
    c_1 &=\frac{\pi}{144}\rho_V^2 P(t_x,t_y,r) K_2/r^2~.
\end{align}
Therefore, two-point correlation function can be expressed as
\begin{align}
&\Pi^{(s)} (t_x,t_y,k) = \int d^3 r e^{i\vec{k} \cdot \vec{r}} K_{ij,kl}(\hat{k})K_{ij,mn}(\hat{k}) 
\langle T_{kl}(t_x, \vec{x}) T_{mn}(t_y, \vec{y}) \rangle^{(s)} \notag \\
&= 
\frac{4\pi^2}{9\beta^3} \rho_V^2\;
\int_{0}^\infty d\Tilde{r} \; P(\tilde{T},\tilde{t}_d,\tilde{r})
\times 
\left[ j_0(\Tilde{k}\Tilde{r})K_0 + \frac{j_1(\Tilde{k}\Tilde{r})}{\Tilde{k}\Tilde{r}}K_1 + \frac{j_2(\Tilde{k}\Tilde{r})}{\Tilde{k}^2\Tilde{r}^2}K_2 \right]~,
\label{eq:Pi_single}
\end{align}
with
\begin{align}
K_0 &= G_0\frac{H^4e^{\tilde{T}-\Tilde{r}/2}}{\beta^4}\frac{1}{\Tilde{r}^3}F_0+n_\mathrm{pbh}\frac{H^3}{\beta^3}\Tilde{r}F''_0~,\\
K_1 &= G_0\frac{H^4e^{\tilde{T}-\Tilde{r}/2}}{\beta^4}\frac{1}{\Tilde{r}^3}F_1+n_\mathrm{pbh}\frac{H^3}{\beta^3}\Tilde{r}F''_1~,\\
K_2 &= G_0\frac{H^4e^{\tilde{T}-\Tilde{r}/2}}{\beta^4}\frac{1}{\Tilde{r}^3}F_2+n_\mathrm{pbh}\frac{H^3}{\beta^3}\Tilde{r}F''_2~,\\
P(\tilde{T},\tilde{t}_d,\tilde{r}) &= \exp \left[-G_0\frac{H^4}{\beta^4}8\pi e^{\tilde{T}}\mathcal{I}(\Tilde{t}_d,\Tilde{r})-n_\mathrm{pbh}\frac{H^3}{\beta^3} I'(\Tilde{r},\Tilde{T},\Tilde{t}_d)\right]~,
\end{align}
where
\begin{align}
    F_0&
= 2(\tilde{r}^2 - \tilde{t}_d^2)^2(\tilde{r}^2 + 6\tilde{r} + 12)~, \\
F_1&
= 2(\tilde{r}^2 - \tilde{t}_d^2)\left[ -\tilde{r}^2(\tilde{r}^3+4\tilde{r}^2 + 12\tilde{r} + 24)  + \tilde{t}_d^2(\tilde{r}^3 + 12\tilde{r}^2 + 60\tilde{r} + 120) \right]~, \\
F_2&
= \frac{1}{2} \left[
\tilde{r}^4(\tilde{r}^4 + 4\tilde{r}^3 + 20\tilde{r}^2 + 72\tilde{r} + 144) \right.
- 2\tilde{t}_d^2\tilde{r}^2(\tilde{r}^4 + 12\tilde{r}^3 + 84\tilde{r}^2 + 360\tilde{r} + 720) \nonumber \\
&\;\;\;\;\;\;\;\;
\left.+ \tilde{t}_d^4(\tilde{r}^4 + 20\tilde{r}^3 + 180\tilde{r}^2 + 840\tilde{r} + 1680) \right]~.
 \\ 
F''_0&= \frac{\left(\tilde{r}^2-4 \tilde{T}^2\right)^2 \left(\tilde{r}^2-\tilde{t}_d^2\right)^2}{16 \tilde{r}^4}~,  \\ 
F''_1 &= \frac{\left(\tilde{r}^2-4 \tilde{T}^2\right) (\tilde{r}-\tilde{t}_d) (\tilde{r}+\tilde{t}_d) \left(3 \tilde{r}^4+\tilde{r}^2 \left(4 \tilde{T}^2+\tilde{t}_d^2\right)-20 \tilde{T}^2 \tilde{t}_d^2\right)}{8 \tilde{r}^4}~,
 \\
F''_2 &= \frac{3 \tilde{r}^8 + 560 \tilde{T}^4 \tilde{t}_d^4 + 2 \tilde{r}^6 (4 \tilde{T}^2 + \tilde{t}_d^2) - 
 120 \tilde{r}^2 \tilde{T}^2 \tilde{t}_d^2 (4 \tilde{T}^2 + \tilde{t}_d^2) + 
 3 \tilde{r}^4 (16 \tilde{T}^4 + 16 \tilde{T}^2 \tilde{t}_d^2 + \tilde{t}_d^4)}{16 \tilde{r}^4}~.
\end{align}
Note that the first terms in functions \(K_0,K_1,K_2,P\) are coincide with standard case~\cite{jinno_gravitational_2017a}, while the second terms represent corrections from PBHs.
Finally single-bubble contribution spectrum \(\Delta^{(s)}\) can be expressed as
\begin{align}
\Delta^{(s)} = \frac{2 \Tilde{k}^3}{ 3}\int_0^\infty d\Tilde{r}\,
\int_0^{\Tilde{r}} d\Tilde{t_d}\,
\int_{\Tilde{r}/2}^\infty d\Tilde{T} \cos(\Tilde{k}\Tilde{t_d})  P(\tilde{T},\tilde{t}_d,\tilde{r})
\times 
\left[ j_0(\Tilde{k}\Tilde{r})K_0 + \frac{j_1(\Tilde{k}\Tilde{r})}{\Tilde{k}\Tilde{r}}K_1 + \frac{j_2(\Tilde{k}\Tilde{r})}{\Tilde{k}^2\Tilde{r}^2}K_2 \right]~.
\label{eq:D_single}
\end{align}

\subsection{Double-Bubble Contribution}

Two conditions are required for two separated bubbles contribute nonvanishing two-point correlation function: (i) No bubble nucleates inside \(V_{xy}\); (ii) Only one bubble nucleates in each region of \(\delta V_x-\l(\delta V_x \cap V_y\r)\) and \(\delta V_y-\l(\delta V_y \cap V_x\r)\). In the following, we use \(\delta V_x^{(y)}\)and \(\delta V_y^{(x)}\) to denote the region \(\delta V_x-\l(\delta V_x \cap V_y\r)\) and \(\delta V_y-\l(\delta V_y \cap V_x\r)\), respectively. Ensemble average gives
\begin{align}
\langle T_{kl}(t_x, \vec{x}) T_{mn}(t_y, \vec{y}) \rangle^{(d)}
&= 
P(t_x,t_y,r) 
\int_{-\infty}^{t_{xy}} dt_{xn}
\Gamma(t_{xn})
\int_{\delta V_x^{(y)} \cap \Sigma_{t_{xn}}} d^3x_n \;
{\mathcal T}^{(d)}_{x,ij}(t_{xn},\vec{x}_n;t_x,\vec{r}) 
\nonumber \\
&\;\;\;\;
\times
\int_{-\infty}^{t_{xy}} dt_{yn}
\Gamma(t_{yn})
\int_{\delta V_y^{(x)} \cap \Sigma_{t_{yn}}} d^3y_n \;
{\mathcal T}^{(d)}_{y,kl}(t_{yn},\vec{y}_n;t_y,\vec{r})~,
\label{eq:Double_TT}
\end{align}
where ${\mathcal T}^{(d)}_{x,ij}$ and ${\mathcal T}^{(d)}_{y,kl}$ 
are the value of the energy-momentum tensor 
generated by the bubble wall nucleated in $\vec{x}_n \in \delta V_x^{(y)} \cap \Sigma_{t_{xn}}$ 
and $\vec{y}_n \in \delta V_y^{(x)} \cap \Sigma_{t_{yn}}$, respectively.
They are given by
\begin{align}
{\mathcal T}^{(d)}_{x,ij}(t_{xn},\vec{x}_n;t_x,\vec{r}) 
= 
\left( \frac{4\pi}{3}r_x(t_{xn})^3 \cdot \rho_0 \cdot \frac{1}{4\pi r_x(t_{xn})^2l_B} \right) 
(n_{x\times})_i (n_{x\times})_j~,
\nonumber \\
{\mathcal T}^{(d)}_{y,kl}(t_{yn},\vec{y}_n;t_y,\vec{r}) 
= 
\left( \frac{4\pi}{3}r_y(t_{yn})^3 \cdot \rho_0 \cdot \frac{1}{4\pi r_y(t_{yn})^2l_B} \right)
(n_{y\times})_i (n_{y\times})_j~.
\end{align}

Since there are no special directions except for $\vec{r}$, ${\mathcal T}_{z,ij}^{(d)}$ ($z = x,y$) can be decomposed as follows 
\begin{align}
\int_{-\infty}^{t_{xy}} dt_{zn} \Gamma(t_{zn})
\int d^3z_n 
\; 
{\mathcal T}^{(d)}_{z,ij}(t_{zn},\vec{z}_n;t_z,\vec{r}) 
= {\mathcal A}^{(d)}_z(t_x,t_y,r) \delta_{ij}
+ {\mathcal B}^{(d)}_z(t_x,t_y,r) \hat{r}_i \hat{r}_j~.
\end{align}
Here ${\mathcal A}_z^{(d)}$ and ${\mathcal B}_z^{(d)}$ depend on 
both $t_x$ and $t_y$ because the integration region for $z_n$ 
is affected by the other points.
After the projection by $K$, only ${\mathcal B}$ component survives:
\begin{align}
K_{ij,kl}(\hat{k})K_{ij,mn}(\hat{k})
\langle T_{kl}(t_x, \vec{x}) T_{mn}(t_y, \vec{y}) \rangle^{(d)} = \frac{1}{2}
P(t_x,t_y,r) 
{\mathcal B}^{(d)}_x(t_x,t_y,r)
{\mathcal B}^{(d)}_y(t_x,t_y,r)
(1 - (\hat{r} \cdot \hat{k})^2)^2~.
\end{align}
Neglecting the contribution from \(t_{zn}<0\), we can calculate \({\cal B}^{(d)}\) as
\begin{align}
{\mathcal B}^{(d)}_x(t_x,t_y,r) &= \frac{\pi \rho_0}{3} \left[G_0\frac{H^4}{\beta^4}C_0(\Tilde{r},\Tilde{T},\Tilde{t_d})+n_\mathrm{pbh}\frac{H^3}{\beta^3}C_1(\Tilde{r},\Tilde{T},\Tilde{t_d})\right]~, \\
{\mathcal B}^{(d)}_y(t_x,t_y,r) &= \frac{\pi \rho_0}{3} \left[G_0\frac{H^4}{\beta^4}C_0(\Tilde{r},\Tilde{T},-\Tilde{t_d})+n_\mathrm{pbh}\frac{H^3}{\beta^3}C_1(\Tilde{r},\Tilde{T},-\Tilde{t_d})\right]~.
\end{align}
where
\begin{align}
C_0(r,T,t_d) &= -\exp \left(T-\frac{r}{2}\right)\frac{(r-t_d) (r+t_d)  \left(  r^3 +2r^2+t_d (  r^2+ 6 r+12)\right)}{2  r^3}~,\\
C_1(r,T,t_d) &= \frac{\left(r^2-4 T^2\right) (r-t_d) (r+t_d) \left(r^2+2 T t_d\right)}{8 r^3}~.
\end{align}
As in the single-bubble case, the angular integration is readily calculated
\begin{align}
\Pi^{(d)}& =
\frac{16\pi^3}{9\beta^3} \rho_0^2
\int_0^\infty d\Tilde{r} \;
P(\tilde{T},\tilde{t}_d,\tilde{r})\Tilde{r}^2 \frac{j_2(\Tilde{k}\Tilde{r})}{\Tilde{k}^2\Tilde{r}^2} \nonumber \\
&\times \left[G_0\frac{H^4}{\beta^4}C_0(\Tilde{r},\Tilde{T},\Tilde{t_d})+n_\mathrm{pbh}\frac{H^3}{\beta^3}C_1(\Tilde{r},\Tilde{T},\Tilde{t_d})\right] \times \left[G_0\frac{H^4}{\beta^4}C_0(\Tilde{r},\Tilde{T},-\Tilde{t_d})+n_\mathrm{pbh}\frac{H^3}{\beta^3}C_1(\Tilde{r},\Tilde{T},-\Tilde{t_d})\right]~.
\label{eq:Pi_double}
\end{align}
Finally, we can get the spectrum for double-bubble contribution,
\begin{align}
\Delta^{(d)} = &\frac{8\pi \Tilde{k}^3}{3} \int_0^\infty d\Tilde{r}\,\int_0^{\Tilde{r}} d\Tilde{t_d}\,
\int_{\Tilde{r}/2}^\infty d\Tilde{T} \cos(\Tilde{k}\Tilde{t_d})  P(\tilde{T},\tilde{t}_d,\tilde{r})\Tilde{r}^2 \frac{j_2(\Tilde{k}\Tilde{r})}{\Tilde{k}^2\Tilde{r}^2}
\nonumber \\
&\times \left[G_0\frac{H^4}{\beta^4}C_0(\Tilde{r},\Tilde{T},\Tilde{t_d})+n_\mathrm{pbh}\frac{H^3}{\beta^3}C_1(\Tilde{r},\Tilde{T},\Tilde{t_d})\right] \times \left[G_0\frac{H^4}{\beta^4}C_0(\Tilde{r},\Tilde{T},-\Tilde{t_d})+n_\mathrm{pbh}\frac{H^3}{\beta^3}C_1(\Tilde{r},\Tilde{T},-\Tilde{t_d})\right]~.
\label{eq:D_double}
\end{align}

\section{SIGW associated with PBHs}
\label{app:sigw}
{
If PBHs are generated from the collapse of large density perturbations, they are unavoidably associated to the emission of induced GWs at second order by the same scalar perturbations due to the intrinsic nonlinear nature of gravity~\cite{sasaki_primordial_2018}. If the primordial curvature power spectrum is a delta function, then there is an one-to-one relation between PBH parameters (\(f_{\rm pbh}\), \(M_{\rm pbh}\)) and induced GW parameters (\(\Omega_{\rm GW}h^2\), \(f_{\rm GW}\))~\cite{Saito:2009jt}. Here we list the main results from delta-peak primordial curvature power spectrum.

Considering primordial power spectrum with \({\cal P }_\Psi = {\cal A}^2\delta(\ln(k/k_p))\), the masses of PBHs can be expressed as
\begin{align}
    M_{\rm pbh} = 10^{20}{\rm g}\left(\frac{k_p}{2\times 10^7{\rm pc}^{-1}}\right)\left(\frac{g_*}{106.75}\right)^{-1/6}~.
\end{align}
And the current mass fraction of PBHs can be expressed as
\begin{align}
    f_{\rm pbh} = 4\times10^{14}\frac{\Psi_c}{{\cal A}}e^{-\Psi_c^2/2{\cal A}^2}\left(\frac{M_{\rm pbh}}{10^{20}{\rm g}}\right)^{-1/2}\left(\frac{g_*}{106.75}\right)^{-1/3}~,
\end{align}
where \(\Psi_c \sim 0.5\)~\cite{Carr:1975qj} is the threshold for gravitational collapse. On the other hand, the GW induced by primordial perturbation can be expressed as
\begin{align}
    \Omega_{\rm GW}h^2 = {\cal N} \frac{2{\cal A}^4}{3}\kappa^{-2}\left(1-\frac{\kappa^2}{4}\right)^2 \overline{I^2(\kappa)}\ \Theta(2-\kappa)~,
\end{align}
where \(\kappa = k/k_p\) and \(\overline{I^2(\kappa)}\) is the kernel function. Here \({\cal N }=1.67\times10^{-5}\). The energy density \(\Omega_{\rm GW}h^2\) has a peak at a frequency \(f_{\rm GW} \equiv k_p/(\sqrt{3}\pi)\) and scales as \(f^2\) for low frequencies. A typical amplitude of the induced GWs at the peak frequency is given by
\begin{align}
    A_{\rm GW} = 6\times10^{-8} \left(\frac{g_*}{106.75}\right)^{-1/3} \left(\frac{{\cal A}^2}{10^{-2}}\right)^2 \left(\frac{\Omega_{\rm rad}h^2}{4\times 10^{-5}}\right)~.
\end{align}
Therefore, combining the result from PBH production and induced GWs emission, we have
\begin{align}
    f_{\rm GW} = 0.03\ {\rm Hz} \left(\frac{M_{\rm pbh}}{10^{-13} M_\odot}\right)^{-1/2}\left(\frac{g_*}{106.75}\right)^{-1/12}~.
\end{align}
Assuming that PBHs with mass \(10^{-14}M_\odot\) constitute all dark matter, which is compatible with PT interpretation of PTA signal if we includes SIGW from PTs, the peak frequency of SIGW from PBH production is \(f_{\rm GW} \approx 0.1\ {\rm Hz}\) and the typical amplitude is \(\Omega_{\rm GW} h^2 \approx 1.5\times 10^{-9}\). In the frequency band of PTA, the GW energy density is \(\Omega_{\rm GW}h^2 \approx 10^{-23}\) at \(f = 10^{-8}\ {\rm Hz}\), which is completely negligible compared to the observed SGWB \(\Omega_{\rm GW}h^2 \sim 10^{-8} \). 
}

\bibliographystyle{refs}
\bibliography{references}

\end{document}